\documentstyle[aps,epsfig]{revtex}

\begin{document}
\author{Anand T.N. Kumar, Florin Rosca, Allan Widom and Paul M. Champion\\
Department of Physics and Center for Interdisciplinary Research on Complex
Systems,\\
Northeastern University, Boston MA 02115}
\title{{\LARGE Investigations of Amplitude and Phase Excitation Profiles in
Femtosecond Coherence Spectroscopy}}
\maketitle
\begin{center}
{\bf Abstract}
\end{center}
We present an effective linear response approach to pump-probe
femtosecond coherence spectroscopy in the well separated pulse
limit. The treatment presented here is based on a displaced and
squeezed state representation for the non-stationary states
induced by an ultrashort pump laser pulse or a chemical reaction.
The subsequent response of the system to a delayed probe pulse is
modeled using closed form non-stationary linear response
functions, valid for a multimode vibronically coupled system at
arbitrary temperature. When pump-probe signals are simulated using
the linear response functions, with the mean nuclear positions and
momenta obtained from a rigorous moment analysis of the pump
induced (doorway) state, the signals are found to be in excellent
agreement with the conventional third order response approach. The
key advantages offered by the moment analysis based linear
response approach include a clear physical interpretation of the
amplitude and phase of oscillatory pump-probe signals, a dramatic
improvement in computation times, a direct connection between
pump-probe signals and equilibrium absorption and dispersion
lineshapes, and the ability to incorporate coherence such as those
created by rapid non-radiative surface crossing. We demonstrate
these aspects using numerical simulations, and also apply the
present approach to the interpretation of experimental amplitude
and phase measurements on reactive and non-reactive samples of the
heme protein Myoglobin. The role played by inhomogeneous
broadening in the observed amplitude and phase profiles is
discussed in detail. We also investigate overtone signals in the
context of reaction driven coherent motion.
\section{\protect\smallskip Introduction}

Femtosecond coherence spectroscopy (FCS) is an ultrafast
pump-probe technique that allows the experimentalist to create and
probe coherent vibrational motions and ultrafast chemical
reactions in real time
\cite{tang86,ruhman87,nelson87,nelson87b,zewail88,Fragnito89,scherer91,pollard92,dexheimer92,zewail93,vos93,dougherty,nelson94,ZHU94,wang94,bradforth95,jonas96,sund95,sund95a,yang99}.
In a typical pump-probe experiment, an ultrashort pump laser pulse
is used to excite the sample of interest. The subsequent
non-stationary response of the medium is monitored by an optically
delayed probe pulse. Owing to the large spectral bandwidth
available in a short laser pulse, it can generate non-stationary
vibrational states in a molecular system as shown in Fig. 1(a).
The subsequent nuclear dynamics modulates the optical response as
detected by the probe pulse. Coherent vibrational motion in the
ground state has been observed in crystalline and liquid phase
systems \cite{ruhman87,dougherty}, and in biological specimens
having short excited state lifetimes such as
bacteriorhodopsin\cite{pollard92,dexheimer92}, and
myoglobin\cite{ZHU94}. For molecules that have long-lived excited
states, however, the excited state coherence is dominant and has
been identified in several dye molecules in
solution\cite{tang86,Fragnito89,yang99}, in small molecules in the
gas phase\cite{zewail88,scherer91,zewail93}, and in photosynthetic
reaction centers\cite{vos93,jonas96}.

Apart from ``field driven'' coherence directly prepared by the
laser fields, vibrational coherence can also be driven by rapid
non-radiative processes. For example, if we consider a third
electronic state $\left| f\right\rangle $ that is coupled
non-radiatively to the photo-excited state $\left| e\right\rangle
$ as in Fig. 1(b), the wave-packet created in the excited state by
the pump can cross over to $\left| f\right\rangle $, leaving a
vibrationally coherent product \cite{wang94}. Fig. 1(b) suggests
the importance of taking a multidimensional view of the problem,
whereby the surface crossing between the reactant excited state
$\left| e\right\rangle $ and the product state, $\left|
f\right\rangle$ along the reaction coordinate $R$ is accompanied
by the creation of a vibrational coherence along the $Q$
coordinate that is coupled to the non-radiative transition. In
earlier work, we have presented expressions for the time dependent
population and first moment evolution of vibrational dynamics
following a Landau-Zener surface crossing\cite{zhu97}. These
expressions are rigorously valid, provided the quantum yield for
the reaction is unity (a condition that holds for NO and CO
photolysis from heme
proteins\cite{schuresco78,gibson86,miller97}).

A common theoretical formulation for pump-probe spectroscopy is
based on the third order susceptibility $\chi ^{(3)}$ formalism,
which provides a unified view of four wave mixing spectroscopies
\cite{hellwarth77,muk82,pollard92b,muk95} with different
combinations of fields, irrespective of whether they are
continuous wave or pulsed. However, the separation of the pump and
probe events is not clear in this formalism, since the pump
induced density matrix is implicitly contained in the third order
response functions. Thus, it is also attractive to treat the pump
and probe processes separately in the well separated pulse (WSP)
limit. For example, the ``doorway/window'' picture has been
developed\cite{muk95} which can be used to represent the pump and
probe events as Wigner phase space wave packets. This readily
enables a semi-classical interpretation of pump-probe
experiments\cite{yan89,yan90,muk90}. Another view of the WSP limit
is based on the effective linear response approach. In this
approach, the pump induced medium is modeled using a
time-dependent linear
susceptibility\cite{fain92,fain93b,fain93,pollard92,pollard92b}.
This has the appealing aspect that a pump-probe experiment is
viewed as the non-stationary extension of steady state absorption
spectroscopy.
\begin{figure}[htbp]
\begin{center}
\mbox{\epsfig{file=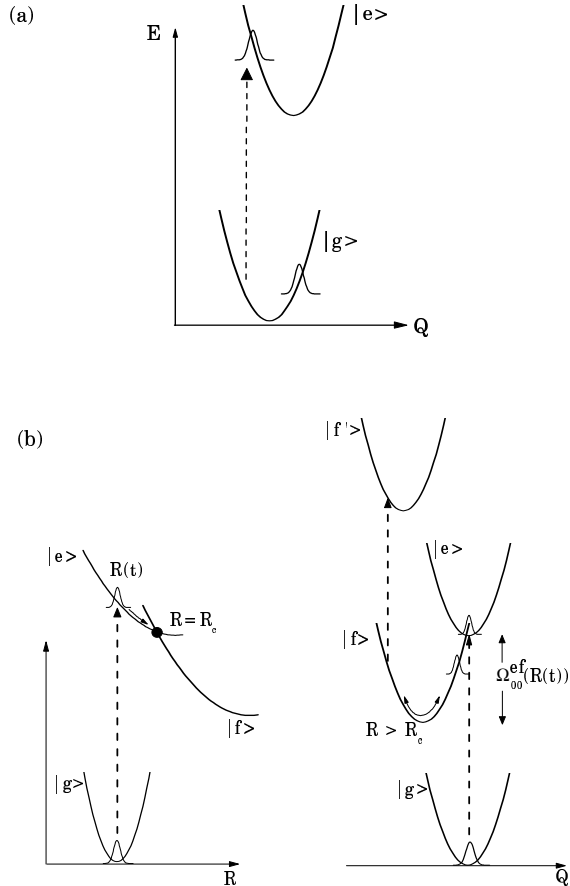,height=120mm}} \caption{(a)
Schematic of a two electronic level system, with ground
($\left|g\right\rangle$) and excited ($\left| e\right\rangle $)
electronic states that are coupled radiatively. (b)
Multidimensional view of vibrational coherence created by a
Landau-Zener surface crossing. Surface crossing along the
R-coordinate between the electronic states $\left|e\right\rangle $
and $\left| f\right\rangle$ leads to vibrational coherence along
the $Q$ coordinate in the product electronic state. The states
$\left| g\right\rangle $ and $\left| f\right\rangle$ are assumed
not be radiatively coupled.}
\end{center}
\end{figure}
While the above theoretical treatments are primarily concerned
with field driven processes, the need to incorporate additional
non-radiative mechanisms into the theory is evident from recent
experimental observations \cite{ZHU94,zhu97,rosca00}. Recent
studies have addressed population transfer during non-adiabatic
transitions in electron transfer systems\cite{coalson94,cho95}.
Numerical wave packet propagation techniques have also been
presented to model coherent dynamics due to non-adiabatic excited
state processes\cite{domcke97,dilthey00}.

The present work is motivated by the need to interpret FCS
experiments of reactive and non-reactive samples of
myoglobin\cite{rosca00}. Here, we make the WSP approximation, and
demonstrate the usefulness and accuracy of the effective linear
response approach, with the non-stationary states represented
using unitary displacement and squeezing operators\cite{mandel95}.
This representation readily leads to rigorous analytic expressions
for the effective linear response functions. The analytic
expressions are physically intuitive and allow for the efficient
calculation of oscillatory amplitude and phase profiles over a
wide range of carrier frequencies. The resulting information is
the FCS analogue of a resonance Raman excitation profile. The
computational ease of this approach eliminates the need for
impulsive or semi-classical
approximations\cite{dilthey00,cina93,cho93a,cho93b,jonas95,smith94,shen99}.
Furthermore, when the linear response functions are used in
combination with a moment analysis of the pump induced (doorway)
state\cite{kumar00} (presented in Appendix A), excellent agreement
is found with the full third order response
approach\cite{pollard92b,muk95}. In addition, coherence driven by
rapid non-radiative processes as in Fig. 1(b) can be readily
incorporated into this formalism. This is illustrated using the
first moments (presented in Appendix B) of a Landau-Zener driven
coherence\cite{zhu97}.

The general outline of the paper is as follows. In Sec.\ II, we
briefly review the third order response approach and its
connection with the effective linear response approach in the well
separated pulse limit. In Sec.\ III, we obtain analytic
expressions for the effective linear response function using a
displaced thermal state representation for the doorway state. We
further consider undamped vibrational motion and derive arbitrary
temperature expressions for the dispersed and open band pump-probe
signals. In Sec.\ IV, we present simulations to demonstrate the
accuracy and physically intuitive aspects of the first moment
based effective linear response approach. Next, we include the
effects of inhomogeneous broadening of the electronic lineshape.
Finally, we consider the application of the effective linear
response approach to reaction driven coherence, and discuss the
magnitude of overtone signals using displaced and squeezed initial
states.

\section{Third order and effective linear response}

Here, we briefly review the connection between the third order and
effective linear response functions in the well separated pulse
limit. In the electric dipole approximation, we ignore the spatial
dependence of the electric fields and write the total electric
field of the incident pulses as ${\bf E}(t)={\bf E}_{a}(t)+{\bf
E}_{b}(t-\tau )$. The subscripts `$a$' and `$b$' refer to the pump
and the probe, and $\tau $ is the delay between the two pulses.
The detected signal in the open band detection scheme, is the
differential probe energy transmitted through the medium
\begin{equation}
S(\tau )=-\int_{-\infty }^{+\infty }dt{\bf E}_{b}(t-\tau )
\cdot (\partial{\bf P}(t,\tau )/\partial t),  \label{1b}
\end{equation}
where ${\bf P}(t,\tau )$ is the material polarization induced by the pump
and probe fields. If we assume that the delay $\tau $ is larger than the
pulse durations, the lowest order in the polarization that is detected in
pump-probe spectroscopy is given by
\begin{equation}
P_{i}(t,\tau )=\int_{-\infty }^{t}dt_{3}\int_{-\infty
}^{t_{3}}dt_{2}\int_{-\infty }^{t_{2}}dt_{1}\chi
_{ijkl}^{(3)}(t,t_{1},t_{2},t_{3})E_{bj}(t_{3}-\tau
)E_{ak}(t_{2})E_{al}(t_{1}).  \label{1}
\end{equation}
Here, $\chi _{ijkl}^{(3)}$ is the tensorial third order
susceptibility. We have written the sequential
contribution\cite{yan89} ignoring the coherent coupling/tunneling
terms\cite{sund95,sund95a} that involve the overlap of the pump
and probe fields. The state of the system in between the pump and
probe pulses is not exposed in the the above expression for the
polarization. If we group the $t_{1}$ and $t_{2}$ integrations in
Eq. (\ref{1}) that involve only the pump interactions, then we are
left with the $t_{3}$ integral over the probe field and we may
rewrite Eq. (\ref{1}) as
\begin{equation} P_{i}(t,\tau )=\int_{-\infty }^{t}dt_{3}\chi
_{ij}^{eff}(t,t_{3})E_{bj}(t_{3}-\tau ),  \label{1a}
\end{equation}
where $\chi _{ij}^{eff}(t,t_{3})$ is the effective linear susceptibility
describing the non-stationary medium created by the pump interaction\cite
{pollard92b,fain92}. Recall for a stationary medium that
$\chi_{ij}^{eff}(t,t_{3})= \chi _{ij}(t-t_{3})$, which describes the
equilibrium response of the system.

Eqs. (\ref{1}) and (\ref{1a}) are formally equivalent expressions for the
polarization induced by the well separated pump and probe pulses.
However, Eq. (\ref{1a}) is more important than a mere
rephrasing of the third order polarization expression in the
well separated limit. The non-stationary response function
$\chi_{ij}^{eff}(t,t_{3})$ can equally well describe coherence induced
by mechanisms other than (and including) the pump field interaction;
e.g. a rapid non-radiative surface crossing. To see what is involved,
we evaluate the induced polarization quantum mechanically
(in the interaction picture\cite{sakurai}) as
\begin{equation}
{\bf P}(t)={\large Tr}\left( \widehat{\rho }(t,\tau )\widehat{{\bf \mu }}%
(t)\right),  \label{11}
\end{equation}
$\widehat{\rho }(t,\tau )$ is density operator describing the
quantum-statistical state of the system, and $\widehat{{\bf \mu
}}$ is the electric dipole moment operator. We write the total
Hamiltonian for the molecule and its interaction with the laser
fields as $\widehat{H}(t)=\widehat{H}_{0}+\widehat{H}_{I}(t),$
where the free Hamiltonian $\widehat{H}_{0}$ and the interaction
Hamiltonian $\widehat{H}_{I}$ have the following form for a
molecular system with two electronic levels:
\begin{equation}
\widehat{H}_0=\pmatrix{\widehat{H}_e+\hbar \Omega_v & 0 \cr 0 &
\widehat{H}_g}; \ \widehat{H}_I(t)=-{\bf \widehat{\mu} \cdot
E}(t)= -\pmatrix{ 0 & {\bf \mu}_{ge}{\bf \cdot E}(t)\cr {\bf
\mu}_{eg}{\bf \cdot E}(t) & 0}, \label{12}
\end{equation}
where $\widehat{H}_{g}$ and $\widehat{H}_{e}=\widehat{H}_{g}+\widehat{V}$,
are, respectively, the Born-Oppenheimer Hamiltonians for the ground
and excited electronic states. $\widehat{V}$ is defined as the difference
potential that specifies the electron nuclear coupling and
$\hbar \Omega _{v}$ is the vertical electronic energy gap at the
equilibrium position of the ground state. With the above definitions,
the third order susceptibility takes the standard form
\begin{equation}
\chi _{ijkl}^{(3)}(t,t_{1},t_{2},t_{3})=(i/\hbar )^3
{\large Tr}\left( \widehat{\mu }_{i}(t)\left[ \widehat{\mu }
_{j}(t_{3}),\left[ \widehat{\mu }_{k}(t_{2}),\left[ \widehat{\mu }
_{l}(t_{1}),\widehat{\rho }(-\infty )\right] \right] \right] \right).
\label{14}
\end{equation}
Here, $\widehat{\rho }(-\infty )$ is the initial density operator
of the system before the pump and the probe pulses and $Tr$
denotes the trace operation. Expansion of the above triple
commutator leads to four non-linear response functions and their
complex conjugates\cite{pollard92,muk95}. On the other hand, if we
substitute Eq. (\ref{14}) in Eq. (\ref{1}), and group the pump
field interactions separately, a comparison of the resulting
expression with Eq. (\ref{1a}) leads to
\begin{equation}
\chi _{ij}^{eff}(t,t_{3})=(i/\hbar )
{\large Tr}\left(\widehat{\mu }_{i}(t)
\left[ \widehat{\mu }_{j}(t_{3}),\delta \widehat{\rho }
(t_{3})\right] \right) = (i/\hbar )
{\large Tr}\left(\delta \widehat{\rho }(t_{3})
\left[ \widehat{\mu }_{i}(t),\widehat{\mu }
_{j}(t_{3})\right] \right) .  \label{14a}
\end{equation}
Here, $\delta \widehat{\rho }(t)$ is the perturbation in the density matrix
due to the pump interaction (also referred to as the density matrix jump%
\cite{fain90} and the doorway function\cite{yan89,muk90}) and is given (in
the interaction picture) by,
\begin{equation}
\delta \widehat{\rho }(t_{3})=\left( \frac{i}{\hbar }\right)
^{2}\int_{-\infty }^{t_{3}}dt_{2}\int_{-\infty }^{t_{2}}dt_{1}\left[
\widehat{\mu }_{k}(t_{2}),\left[ \widehat{\mu }_{l}(t_{1}),\widehat{\rho }%
(-\infty )\right] \right] E_{ak}(t_{2})E_{al}(t_{1}).  \label{15a}
\end{equation}
In arriving at the second equality in Eq. (\ref{14a}), we have used the
invariance of the trace under permutation. In the WSP limit, where the
overlap between the pump and probe pulse is negligible, we can extend
the upper limit $t_{3}$ in Eq. (\ref{15a}) to
infinity. The interaction picture density matrix is then time independent.

Eq. (\ref{14a}) expresses the effective linear response as the average value
of the standard linear response Green's function\cite{martin68}
with respect to the pump induced (non-stationary) density matrix
$\delta \widehat{\rho }$. Note that $\delta \widehat{\rho }$
does not necessarily have to be induced by the pump fields as in
Eq. (\ref{15a}). It could, for example, describe the perturbation
in the state of the system due to a non-radiative interaction that follows
the pump excitation to a dissociative state. It might even be a pump induced
state with higher order pump interactions included as would be required
for very strong pump fields\cite{banin94}. The only assumptions are
that the probe field interrogates the non-stationary medium after
the coherence has been created, and the probe interaction with the
medium is linear.

We express the full perturbed density matrix $\delta \widehat{\rho }$
in terms of the ground and excited electronic state nuclear
sub-matrices. The point is that the second order density matrix in
Eq. (\ref{15a}) has no electronic coherence due to the even number of
dipole interactions. Hence
\begin{equation}
\delta \widehat{\rho }=\delta \widehat{\rho }_{e}\left| e\right>
\left< e\right| +\delta \widehat{\rho }_{g}\left| g\right> \left<
g\right| = \pmatrix{ \delta \widehat{\rho }_{e} & 0 \cr 0 & \delta
\widehat{\rho }_{g}}. \label{2}
\end{equation}
Correspondingly, the effective linear susceptibility
Eq. (\ref{14a}) can be decomposed into a ground and excited state
linear response function
\begin{equation}
\chi _{ij}^{_{eff}}(t,t_{3})=\chi _{ij}^{(g)}(t,t_{3})+\chi
_{ij}^{(e)}(t,t_{3}),  \label{7}
\end{equation}
where
\begin{equation}
\chi _{ij}^{(e,g)}(t,t_{3})=(i/\hbar) Tr\left[ \delta
\widehat{\rho }_{e,g}\left\langle e,g\right| \left[ \widehat{\mu
}_{i}(t), \widehat{\mu }_{j}(t_{3})\right] \left| e,g\right\rangle
\right]. \label{8}
\end{equation}

In general, the pulse induced nuclear density matrices $\delta
\widehat{\rho }_{g}$ and $\delta \widehat{\rho }_{e}$ contain
vibrational coherence, i.e. off-diagonal elements in the phonon
number state representation.  This coherence translates into time
dependent wave-packets in a semi-classical phase space $(Q,P)$
Wigner representation of the density matrix
\cite{muk90,tanimura93,dunn95,muk95}. On the other hand, the
highly localized (in $Q$ and $P$) nature of the impulsively
excited non-stationary states suggests that we calculate their
moments using
\begin{equation}
\overline{X}_{e,g}(t)=Tr\left[\left< e,g \right|
\widehat{X}(t)\left| e,g\right> \delta \widehat{\rho
}_{e,g}\right]. \label{3}
\end{equation}
With $\widehat{a}^\dagger (\widehat{a})$ as the creation
(destruction) operator for the phonon mode, $\widehat{X}$
represents the dimensionless quadrature operators
$\widehat{Q}=(\widehat{a}^\dagger + \widehat{a})/\sqrt{2}$,
$\widehat{P}=i(\widehat{a}^\dagger - \widehat{a})/\sqrt{2}$ or
their higher powers. The time $t$ is larger than the pulse
duration. Let the initial ($t=0$) values of the first moments of
the position $\overline{Q}_{s}(0)$ and momentum
$\overline{P}_{s}(0)$, for (say) $s=e$ or $s=g$ be denoted by
$(Q_{s0},P_{s0})$. These uniquely determine the subsequent first
moment dynamics $\overline{Q}_{s}(t)$. Since $Q_{s0}$ and $P_{s0}$
denote shifts from thermal equilibrium, we may represent the pump
induced nuclear density matrix for the electronic state $s$
\begin{equation}
\delta \widehat{\rho }_{s}=\widehat{D}(\lambda _{s})\widehat{\rho }_{T}^{(s)}%
\widehat{D}^{\dagger }(\lambda _{s}),  \label{4}
\end{equation}
where $\widehat{\rho }_{T}^{(s)}$ is the equilibrium thermal density
matrix corresponding to the electronic level $s$.
\begin{equation}
\widehat{\rho }_{T}^{(s)}= Z_{s}^{-1}\exp \left(
-\widehat{H}_{s}/k_{B}T\right).  \label{4a}
\end{equation}
$\widehat{D}(\lambda _{s})$ is the quantum mechanical displacement
operator\cite{foot1} defined as
\begin{equation}
\widehat{D}(\lambda _{s})=\exp \left( \lambda _{s}\widehat{a}^{\dagger
}-\lambda _{s}^{*}\widehat{a}\right).  \label{5}
\end{equation}
Here $\lambda _{s}=\left( Q_{s0}+iP_{s0}\right) /\sqrt{2}$ is the
initial displacement (in phase space) of the coherent state
induced on the potential surface $s$. When Eq. (\ref{4}) is
substituted into Eq. (\ref{8}), the response functions $\chi
^{(e,g)}$ are readily calculated as we show below. Furthermore,
Eq. (\ref{4}) offers a general scheme to represent the
non-stationary density matrix on any given electronic level with
only a knowledge of the first moment $\lambda _{s}$. The effective
linear response function is thus not restricted to coherences
driven by a second order pump interaction. This is the point of
departure of the present work from earlier
treatments\cite{fain92,fain93b,pollard92,pollard92b} which
generally utilize the second order pump induced density matrix Eq.
(\ref{15a}), and are thus formally identical to the $\chi^{(3)}$
approach. This development can be extended to include the higher
moments of $\widehat{Q}$ and $\widehat{P}$, but for harmonic
potentials we expect the higher moments to play a less significant
role in the overall dynamics. Second moment changes are
incorporated in a manner analogous to Eq. (\ref{4}) in Appendix D.

\subsection{Detection schemes}

The two common experimental detection schemes are the open band and the
dispersed probe configurations. For simplicity, we assume that the medium
is isotropic and take the pulse fields to be scalar quantities.
We then write the susceptibilities without tensor subscripts.
The measured open band signal is given by Eq. (\ref{1b}).
For an almost monochromatic laser pulse, we can write the electric
field and the induced polarization as:
\begin{equation}
E_{b}(t-\tau )=\Re e
\left\{ {\cal E}(t-\tau )e^{-i\omega _{c}(t-\tau )}\right\}
\ {\rm and}\
P(t,\tau )=
\Re e
\left\{ {\cal P}(t,\tau )e^{-i\omega _{c}(t-\tau )}\right\} ,  \label{18}
\end{equation}
where $\omega _{c}$ is the carrier frequency, and ${\cal E}(t-\tau
)$ and ${\cal P}(t,\tau )$ are slowly varying envelope functions.
Typically, the envelope function for the probe field is a real
Gaussian centered at time $\tau $; ${\cal E}(t-\tau
)=E_{0}G(t-\tau )$, where $E_{0}$ is the field strength of the
probe pulse. The envelope function for the polarization can be
obtained by using the definitions in Eq. (\ref{18}) and Eq.
(\ref{1a}). After making a change of variables $s=t-t_{3}$, we
find
\begin{equation}
{\cal P}(t,\tau )=\int_{0}^{\infty }dse^{i\omega _{c}s}\chi ^{eff}(t,t-s)
{\cal E}(t-s-\tau ).  \label{18a}
\end{equation}
Employing Eq. (\ref{18}), Eq. (\ref{1b}), and
the rotating wave approximation (RWA) in which highly oscillating
non-resonant terms are ignored, we get
\begin{equation}
S(\tau )=-(\omega _{c}/2)\Im m
\int_{-\infty }^{\infty }dt{\cal E}^{*}(t-\tau ){\cal P}(t,\tau ).
\label{21}
\end{equation}
Since for the present case ${\cal E}(t)$ is real, the measured
(dichroic) signal is directly related to $\Im m \{{\cal P}(t,\tau
)\}$. If ${\cal E}(t)$ were made imaginary as in heterodyne
techniques\cite{zeigler94,cho93a}, then the resulting
(birefringent) signal is related $\Re e \{{\cal P}(t,\tau )\}$.

In the dispersed probe detection scheme, the measured quantity is
spectral density $\widetilde{S}(\omega ,\tau )$ of the transmitted energy
through the sample defined by
\begin{equation}
S(\tau )=\int_{0}^{\infty }\widetilde{S}(\omega ,\tau )d\omega .
\label{10}
\end{equation}
In the RWA,
\begin{equation}
\widetilde{S}(\omega ,\tau )=-(\omega /4\pi )\Im m
\left\{ e^{-i(\omega -\omega _{c})\tau }\widetilde{{\cal E}}^{*}(\omega
-\omega _{c})\widetilde{{\cal P}}(\omega -\omega _{c},\tau )\right\} ,
\label{22}
\end{equation}
where $\widetilde{{\cal E}}(\omega )$ and $\widetilde{{\cal P}}(\omega )$
denote the Fourier transforms of the envelope functions, ${\cal E}(t)$
and ${\cal P}(t,\tau )$.

\section{Evaluation of the response functions}

In this section, we evaluate both the third order and the
effective linear response functions. Consider the ground state
response function $\chi _{g}(t,t_{3})$ given by Eq. (\ref{8})
without the tensor subscripts. We make the Condon approximation
and ignore the coordinate dependence of the dipole moment
operator. If we then let $\widehat{\mu }=\mu _{ge}\left|
g\right\rangle \left\langle e\right| +\mu _{eg}\left|
e\right\rangle
\left\langle g\right| ,$ and use the interaction picture time evolution $%
\widehat{\mu }(t)=e^{iH_{0}t/\hbar }\widehat{\mu }e^{-iH_{0}t/\hbar }$, we
can write,
\begin{equation}
\chi _{g}(t,t_{3})=\left( i\left| \mu _{ge}\right| ^{2}/\hbar \right) \left[
C_{g}(t,t_{3})-C_{g}^{*}(t,t_{3})\right] ,  \label{23}
\end{equation}
where we have defined the two-time correlation function for the ground state
response:
\begin{equation}
C_{g}(t,t_{3})=e^{-i\Omega _{v}(t-t_{3})}Tr\left[ \delta \widehat{\rho }%
_{g}\exp \left( -%
{\displaystyle {i \over \hbar }}%
\int_{t_{3}}^{t}ds\widehat{V}(s)\right) _{+}\right] .  \label{25}
\end{equation}
Here, the subscript $+$ denotes time ordering and $\widehat{V}(s)$ evolves in
time $s$ via $\widehat{H}_{g}$. The density matrix
$\delta \widehat{\rho }_{g}$ is obtained from Eq. (\ref{15a}) as $%
\left\langle g\right| \delta \widehat{\rho }\left| g\right\rangle $,
assuming $\widehat{\rho }(-\infty )=\widehat{\rho }_{T}^{(g)}\left|
g\right\rangle \left\langle g\right| $:
\begin{eqnarray}
\delta \widehat{\rho }_{g} &=&-%
{\displaystyle {\left| \mu _{ge}\right| ^{2} \over \hbar ^{2}}}%
\int_{-\infty }^{\infty }dt_{2}\int_{-\infty
}^{t_{2}}dt_{1}E_{a}(t_{2})E_{a}(t_{1})  \nonumber \\
&&\times \left\{ e^{-i\Omega _{v}(t_{2}-t_{1})}\exp \left( -%
{\displaystyle {i \over \hbar }}%
\int_{t_{1}}^{t_{2}}ds\widehat{V}(s)\right) _{+}\widehat{\rho }%
_{T}^{(g)}+h.c.\right\} ,  \label{26}
\end{eqnarray}
where $h.c.$ denotes Hermitian conjugate.

For the excited state response function, we similarly write
\begin{equation}
\chi _{e}(t,t_{3})=\left( i\left| \mu _{ge}\right| ^{2}/\hbar \right) \left[
C_{e}(t,t_{3})-C_{e}^{*}(t,t_{3})\right] ,  \label{31}
\end{equation}
where $C_{e}$ is the two-time correlation function for the excited
state response:
\begin{equation}
C_{e}(t,t_{3})= e^{i\Omega _{v}(t-t_{3})}Tr\left[ \delta
\widehat{\rho }_{e} \exp \left({\displaystyle {i \over \hbar }}
\int_{0}^{t}ds\widehat{V}(s)\right) _{-}\exp \left( -
{\displaystyle {i \over \hbar }}
\int_{0}^{t_{3}}ds\widehat{V}(s)\right) _{+}\right],  \label{32a}
\end{equation}
with the subscript $-$ denoting anti-time ordering. The excited
state density matrix $\delta \widehat{\rho }_{e}$ is obtained as
$\left\langle e\right| \delta \widehat{\rho }\left| e\right\rangle
$:
\begin{eqnarray}
\delta \widehat{\rho }_{e} &=&%
{\displaystyle {\left| \mu _{ge}\right| ^{2} \over \hbar ^{2}}}%
\int_{-\infty }^{\infty }dt_{2}\int_{-\infty
}^{t_{2}}dt_{1}E_{a}(t_{1})E_{a}(t_{2})  \nonumber \\
&&\times \left\{ e^{i\Omega _{v}s}\exp \left(
{\displaystyle {i \over \hbar }}%
\int_{0}^{t_{2}}ds^{\prime }\widehat{V}(s^{\prime })\right) _{-}\widehat{%
\rho }_{T}^{(g)}\exp \left( -%
{\displaystyle {i \over \hbar }}%
\int_{0}^{t_{1}}ds^{\prime }\widehat{V}(s^{\prime })\right)
_{+}+h.c.\right\} .  \label{33}
\end{eqnarray}

The above expressions are valid for a two level system with arbitrary
difference potentials. In what follows, we take $\widehat{V}=-(\hbar
/m\omega _{0})^{1/2}f\widehat{Q}$ $=-\left( \hbar \omega _{0}\Delta \right)
\widehat{Q}$, with dimensionless $\widehat{Q}$ and relative displacement
$\Delta $. The electron-nuclear coupling force is expressed as
$f=\Delta (m\omega _{0}^{3}\hbar )^{1/2}$, where $m$ and $\omega _{0}$
are, respectively, the reduced mass and frequency of the mode.

If the trace in Eqs. (\ref{25}) and (\ref{32a}) were evaluated
with respect to the equilibrium thermal density matrices
$\widehat{\rho }_{T}^{(e,g)}$ rather than $\delta \widehat{\rho
}_{e,g}$, then the correlation functions,
$C_{e,g}(t,t_{3})=K_{e,g}(t-t_3)$ would be the equilibrium
absorption and emission correlators\cite{page91}; i.e.
\begin{equation}
K_{g}(s)=e^{-i\Omega _{00}s-\Gamma _{e}\left| s\right| }e^{-g(s)};
\ \ K_{e}(s)=e^{i\Omega _{00}s-\Gamma _{e}\left| s\right| }e^{-g(s)},
\label{41}
\end{equation}
where we have introduced homogeneous dephasing through the electronic
damping constant $\Gamma _{e}$. For linearly displaced and undamped
oscillators, $g(s)$ is given by
\begin{equation}
g(s)=\left( \Delta ^{2}/2\right) \left[ (2\overline{n}_{T}+1)\left( 1-\cos
(\omega _{0}s)\right) +i\sin (\omega _{0}s)\right] ,  \label{38}
\end{equation}
where $2\overline{n}_{T}+1=\coth \left( {\hbar \omega _{0}/}2k_{B}T\right) $.
Exact expressions for the damped harmonic oscillator\cite{gu94} can be
easily incorporated into the present development. The half-Fourier
transforms of $K_{e,g}(t)$ determine complex lineshape functions,
\begin{equation}
\Phi (\omega )=i\int_{0}^{\infty }dse^{i\omega s}K_{g}(s), \ \
{\rm  and}\ \ \Theta(\omega )= i\int_{0}^{\infty }dse^{i\omega
s}K_{e}^{*}(s), \label{41d}
\end{equation}
whose imaginary parts are directly related to the absorption and emission
cross-section: $\sigma _{A}(\omega )\propto \omega \Phi
_{I}(\omega )$ and $\sigma _{F}(\omega )\propto \omega ^{3}\Theta
_{I}(\omega )$.

\subsection{Full third-order response}

Analytic expressions have been derived for the third order
susceptibility $\chi ^{(3)}$ for a two level system coupled to a
multimode set of linearly displaced harmonic oscillators, and
expressed in terms of non-linear response functions $R_{j}$
($j=1..4$) \cite{muk82,muk95}. For completeness, these are listed
in Appendix C. When the second order pump induced density matrices
in Eqs. (\ref{26}) and (\ref{33}) are substituted in Eqs.
({\ref{25}) and (\ref{32a}), the ground and excited state
correlation functions $C_{g,e}$ are related to the non-linear
response functions as expressed in Eqs. (\ref{27}) and (\ref{34}).

The expressions of Appendix C cast the effective linear response
functions in terms of the conventional non-linear response
functions. The two approaches are entirely equivalent at this
stage. The key point from Eqs. (\ref{27}) and (\ref{34}) is that
the ground and excited state correlation functions
$C_{g,e}(t,t_{3})$ involve a double integration over the pump
electric fields. In the next section, we will show that the
displaced state representation for the non-stationary density
matrix Eq. (\ref{4}) directly leads to analytic expressions for
$C_{g,e}(t,t_{3})$.

\subsection{Effective linear response}

\subsubsection{Displaced thermal state}
Following Eq. (\ref{4}), we represent the pump induced
non-stationary state as
\begin{equation}
\delta \widehat{\rho }=\widehat{D}(\lambda _{g})\widehat{\rho }_{T}^{(g)}%
\widehat{D}^{\dagger }(\lambda _{g})\left| g\right\rangle \left\langle
g\right| +\widehat{D}(\lambda _{e})\widehat{\rho }_{T}^{(e)}\widehat{D}%
^{\dagger }(\lambda _{e})\left| e\right\rangle \left\langle e\right| .
\label{42}
\end{equation}
This representation can be used to calculate the effective linear
response in a straightforward manner\cite{foot2}. Substituting the
ground state part of Eq. (\ref{42}) into the correlation function
in Eq. (\ref{25}) and using the permutation invariance of the
trace and the unitarity of the displacement operator, we obtain
\begin{equation}
C_{g}{\large (}t{\large ,}t_{3})=e^{-i\Omega _{v}(t-t_{3})}Tr\left[ \widehat{%
\rho }_{T}^{(g)}\exp \left( i\omega _{0}\Delta \int_{t_{3}}^{t}ds\widehat{D}%
^{\dagger }(\lambda _{g})\widehat{Q}(s)\widehat{D}^{\dagger }(\lambda
_{g})\right) _{+}\right] .  \label{45}
\end{equation}
The effect of the displacement operator is to shift the position operator by
a time dependent classical function (c-number), namely the mean pump induced
displacement\cite{mandel95}:
\begin{equation}
\widehat{D}^{\dagger }(\lambda _{g})\widehat{Q}(s)
\widehat{D}(\lambda _{g})=
\widehat{Q}(s)+\overline{Q}_{g}(s),  \label{46}
\end{equation}
where $\overline{Q}_{g}(s)=\Re e\{\lambda _{g}e^{-i\omega
_{0}s}\}$ is the time dependent mean position of the oscillator in
the ground state. If we substitute the above expression into Eq.
(\ref{45}), then the c-number $\overline{Q}_{g}(s)$ can be removed
outside the time ordering and the thermal average simply reduces
to the equilibrium absorption correlator $K_{g}(t-t_{3})$ in Eq.
(\ref{41}). Using Eq. (\ref{47}) of Appendix A for the first
moment dynamics, we find
\begin{equation}
C_{g}\left(t,t_{3}\right)=K_{g}(t-t_{3})\exp \left( i\omega
_{0}A_{g}\Delta \int_{t_{3}}^{t}dse^{-\gamma \left| s\right| }\cos
(\omega _{0}s+\varphi _{g})\right),  \label{49}
\end{equation}
where $A_{g}=\left| \lambda _{g}\right|$ and $\varphi _{g}=\arg
(\lambda_{g}^{*})=-\tan ^{-1}\left( P_{g0}/Q_{g0}\right)$ are the
amplitude and phase for the coherent wave-packet motion in the
ground electronic state. The above expression has an intuitively
appealing form: the ground state correlation function for the pump
induced non-stationary medium is expressed as a modulation of the
standard equilibrium (linear) absorption correlator\cite{page91}
$K_{g}(t-t_3)$ by the first moment dynamics of the ground state
wave-packet motion. It is clear that the corresponding
non-stationary ground state response function $\chi
_{g}(t,t_{3})$, Eq. (\ref{23}), would translate in the frequency
domain into dynamic absorption and dispersion
lineshapes\cite{fain92,pollard92,pollard92b}, which determine the
final probe response to the non-stationary medium. After
performing the integral, Eq. (\ref{49}) reads
\begin{eqnarray}
C_{g}(t,t_{3}) &=&K_{g}(t-t_{3})\exp {\Big [} iA_g\Delta /\omega _{0}%
{\Big \{ }e^{-\gamma t}\left( \omega _{v}\sin (\omega
_{v}t+\varphi _{g})-\gamma \cos (\omega _{v}t+\varphi _{g})\right)
\nonumber \\ &&-e^{-\gamma t_{3}}\left( \omega _{v}\sin (\omega
_{v}t_{3}+\varphi _{g})-\gamma \cos (\omega _{v}t_{3}+\varphi
_{g})\right) {\Big \}} {\Big ]}. \label{50}
\end{eqnarray}
It is interesting to note that the strength of the first moment modulation
appears through the product of the initial displacement and the optical
coupling, $A_{g}\Delta $.

For the excited state correlation function, we similarly find
\begin{eqnarray}
C_{e}(t,t_{3}) &=&K_{e}(t-t_{3})\exp {\Big [}-iA_{e}\Delta /\omega _{0}%
{\Big \{}e^{-\gamma t}\left( \omega _{v}\sin (\omega _{v}t)-\gamma
\cos (\omega _{v}t)\right)  \nonumber \\ &&-e^{-\gamma
t_{3}}\left( \omega _{v}\sin (\omega _{v}t_{3})-\gamma \cos
(\omega _{v}t_{3})\right) {\Big \} } {\Big ] }.  \label{54}
\end{eqnarray}
Thus, the excited state non-equilibrium correlation function is
expressed as the modulation of the equilibrium (fluorescence)
correlation function by the excited state wave-packet.

The analytic expressions in Eqs. (\ref{50}) and (\ref{54}) together
with the first moments presented in Eqs. (\ref{41a}), (\ref{41b}) and
(\ref{47a}), effectively replace the non-linear response expressions derived
in Eqs. (\ref{27}) and (\ref{34}). The generalization of the above single
mode results to the multimode case is straightforward. It easy to show that
the multimode expressions for $C_{g}$ and $C_{e}$, factor into a product
of single mode correlation functions. The results derived using the first
moments are approximate, since the higher moments of the wave-packet
motion (which are induced by the pump pulse\cite{kumar00}) are neglected.
Nevertheless, they provide a fully quantum mechanical description of the
probe response due to modulation by coherent nuclear dynamics. Moreover,
we expect that for a harmonic system, the first moment dynamics will
constitute a major part of the wave-packet motion detected by the probe.
This expectation is verified when we present simulations comparing the
first moment based linear response approach with the full third order
approach.

\subsubsection{Analytic expressions for pump-probe signal: CARS and CSRS
responses}

An advantage of the analytic expressions in Eqs. (\ref{50}) and (\ref{54}),
is that for the special case of undamped nuclear motion, they can be
expanded in a Fourier-Bessel series, from which the polarization
Eq. (\ref{18a}) can be evaluated in closed form. This yields an analytic
expression for the dispersed probe signal Eq. (\ref{22}) that provides
insight into the origin of both resonant and non-resonant pump-probe
signals. If the series expansion
\begin{equation}
e^{i\varrho sin(\theta )}=\sum_{m=-\infty }^{\infty }\exp(im\theta
) J_{m}(\varrho ), \label{55}
\end{equation}
is placed into Eq. (\ref{50}), we find for $\gamma =0$ that
\begin{equation}
C_{g}{\large (}t{\large ,}t_{3})=K_{g}(t-t_{3})\sum_{k=-\infty }^{\infty
}e^{ik(\omega _{0}t_{3}+\varphi _{g})}\sum_{n=-\infty }^{\infty
}J_{n}(A_{g}\Delta )J_{n-k}(A_{g}\Delta )e^{in\omega _{0}(t-t_{3})}.
\label{56a}
\end{equation}
We put this expression into Eqs. (\ref{23}) and (\ref{18a}) to find the
polarization within the RWA (which neglects the contribution of
$C_{g}^{*}$). With Gaussian pulses,
$\widetilde{{\cal E}}(\omega )=E_{0}\widetilde{G}(\omega )$,
where $\widetilde{G}(\omega )\,$ is a Gaussian spectral
function. The ground state contribution to the frequency dispersed
pump-probe signal is finally obtained:
\begin{mathletters}
\begin{eqnarray}
\widetilde{S}_{g}(\omega ,\tau ) &=&\sum_{k=0}^{\infty }{\cal C}_{k}(\omega
)\cos \left( k\left( \omega _{0}\tau +\varphi _{g}\right) \right) +{\cal S}
_{k}(\omega )\sin \left( k\left( \omega _{0}\tau +\varphi _{g}\right) \right)
\label{65x} \\
&=&\sum_{k=0}^{\infty }\widetilde{A}_{k}(\omega )\cos \left( k\omega
_{0}\tau +\widetilde{\varphi }_{k}(\omega )\right) ,  \label{65}
\end{eqnarray}
where the amplitude and phase of the $k$'th overtone are
\end{mathletters}
\begin{mathletters}
\begin{eqnarray}
\widetilde{A}_{k}(\omega ) &=&\left[ {\cal C}_{k}^{2}(\omega )+{\cal S}%
_{k}^{2}(\omega )\right] ^{1/2},  \label{65c} \\
\widetilde{\phi }_{k}(\omega ) &=&k\varphi _{g}-\tan ^{-1}\left[ {\cal S}%
_{k}(\omega )/{\cal C}_{k}(\omega )\right], \label{65d}
\end{eqnarray}
\end{mathletters}
and where ${\cal C}_{k}(\omega )$ and ${\cal S}_{k}(\omega )$ are the
quadrature amplitudes given by,
\begin{mathletters}
\begin{eqnarray}
{\cal C}_{k}(\omega ) &=&\omega M\left[ \widetilde{G}_{p}(\omega
_{d},k\omega _{0})\Phi _{I}^{(k)}(\omega )+\widetilde{G}_{p}(\omega
_{d},-k\omega _{0})\Phi _{I}^{(-k)}(\omega )\right],  \label{65a} \\
{\cal S}_{k}(\omega ) &=&\omega M\left[ \widetilde{G}_{p}(\omega
_{d},k\omega _{0})\Phi _{R}^{(k)}(\omega )-\widetilde{G}_{p}(\omega
_{d},-k\omega _{0})\Phi _{R}^{(-k)}(\omega )\right] .  \label{65b}
\end{eqnarray}
\end{mathletters}
In the above expressions, we have defined the constant
$M=-(| \mu_{ge}| ^{2}E_{0}^{2}/4\pi \hbar) $ for convenience,
and introduced the product spectral function of the probe pulse:
\begin{equation}
\widetilde{G}_{p}(\omega _{d},k\omega _{0})=\widetilde{G}(\omega _{d})%
\widetilde{G}(\omega _{d}+k\omega _{0}),  \label{65cc}
\end{equation}
where $\omega _{d}=\omega -\omega _{c}$ is the detuning frequency and
the absorptive and dispersive basis functions are defined as
\begin{mathletters}
\begin{eqnarray}
\Phi _{I}^{(k)}(\omega ) &=&\sum_{n=-\infty }^{\infty }J_{n}(A_{g}\Delta
)J_{n-k}(A_{g}\Delta )\Phi _{I}(\omega +n\omega _{0}),  \label{66} \\
\Phi _{R}^{(k)}(\omega ) &=&\sum_{n=-\infty }^{\infty }J_{n}(A_{g}\Delta
)J_{n-k}(A_{g}\Delta )\Phi _{R}(\omega +n\omega _{0}).  \label{67}
\end{eqnarray}
\end{mathletters}
The basis functions depend only on $A_g$, $\Delta $ and the equilibrium
lineshape functions and do not depend on the properties of the probe
pulse. Note the symmetry conditions
$\Phi_{I}^{(-k)}(\omega )=\Phi _{I}^{(k)}(\omega -k\omega _{0})$
and $\Phi_{R}^{(-k)}(\omega )=\Phi _{R}^{(k)}(\omega -k\omega _{0})$.
Eqs. (38-42) are rigorous expressions for the dispersed probe signal
for a single undamped mode coupled to a two electronic level system.
They are valid for arbitrary temperature, detection frequency and pulse
width. The simple and direct connection between pump-probe signals
and the equilibrium lineshape functions is exposed by these results.
Equilibrium lineshapes thus enter the calculation at two separate
stages: firstly in the expressions for the pump induced first moment
amplitude and phase in Eqs. (\ref{41a}) and (\ref{41b}), and secondly
in the final probe detected signal in Eqs. (\ref{66}-\ref{67}).

It is clear from Eqs. (38) that the detuning dependence of the
dispersed probe signal is determined by the amplitudes ${\cal
C}_{k}(\omega ) $ and ${\cal S}_{k}(\omega )$, which are in turn
related to the product of the spectral function
$\widetilde{G}_{p}$ and the displaced lineshape functions $\Phi
_{I,R}^{(k)}$. In studying the detuning dependence of
$\widetilde{S}_{g}(\omega ,\tau )$, it is interesting to look at
two opposite limits. When the electronic dephasing time is much
longer than the pulse durations and the vibrational periods, the
basis functions $\Phi_{I}^{(k)}(\omega )$ and $\Phi
_{R}^{(k)}(\omega )$ consist of a well resolved Franck-Condon
progression that acts like a filter in Eqs. (\ref{65a}) and
(\ref{65b}) and determines the detuning dependence. In the
opposite limit, the electronic dephasing time is much shorter than
the pulse durations and vibrational periods. In this case, the
lineshape functions are broad, exhibiting a much slower variation
with respect to $\omega $ than the pulse envelope spectral
function which then acts like a filter in Eqs. (\ref{65a}) and
(\ref{65b}). The dispersed signal then consists of a superposition
of red and blue shifted field envelope functions
$\widetilde{G}_{p}(\omega _{d},\pm k\omega _{0})$. The
superposition weighting depends on the amplitude of the initial
pump induced displacement and the displaced equilibrium lineshape
functions. Thus, resonances occur at frequencies $\omega
\approx\omega _{c}\mp k\omega_{0}/2$ which correspond to the peaks
of $\widetilde{G}_{p}(\omega_{d},\pm k\omega _{0})$. These
resonances can be identified with the well known coherent Stokes
Raman scattering (CSRS) and coherent anti-Stokes Raman scattering
(CARS) signals\cite{druet81,walsh89}.

The delay dependence of the dispersed signal is composed of
oscillations at all harmonics of the fundamental frequency $\omega
_{0}$. The amplitude and the phase of the $k^{th}$ harmonic is
given by $\widetilde{A}_{k}(\omega )$ and $\widetilde{\phi
}_{k}(\omega )$. It is clear from Eq. (\ref{65d}) that the
dispersed probe signal phase $\widetilde{\phi }_{k}(\omega )$ is
not simply related to the initial phase $\varphi _{g}$ of the
coherent motion because of the additional frequency dependent
functions ${\cal C}_{k}(\omega )$ and ${\cal S}_{k}(\omega )$.
However, it can be easily shown that when we integrate
$\widetilde{S}_{g}(\omega ,\tau )$ as specified in Eq. (\ref{10}),
the integral over ${\cal S}_{k}(\omega )$ is small due to
approximately cancelling contributions from the Stokes and
anti-Stokes shifted components in Eq. (\ref{65b}). We then get for
the ground state contribution to the open band signal
\begin{equation}
S_{g}(\tau )=\sum\limits_{k=0}^{\infty}A_{sk}\cos \left( k\omega
_{0}\tau +\phi _{sk}\right) ,  \label{67b}
\end{equation}
where the amplitude and phase of the $k$'th overtone are given by
\begin{mathletters}
\begin{eqnarray}
A_{sk} &=&\int_{0}^{\infty }d\omega \left| {\cal C}_{k}(\omega )\right| ,
\label{67cc} \\
\phi _{sk} &=&k\varphi _{g}-[0\text{ or }\pi ].  \label{68cc}
\end{eqnarray}
\end{mathletters}
Thus, the phase of the fundamental open band signal $\phi_{s1}$
determines the initial phase $\varphi_g $ of the nuclear motion to
within an additive constant of $\pi $. The additive constant phase
arises from the fact that ${\cal C}_{k}(\omega )$ can have a
positive or negative value depending on whether the initial
transmission of the probe pulse is increased or decreased. It is
easily shown that the open band signal will be of the form in Eq.
(\ref{67b}) even when several vibrational modes are active. These
results demonstrate that the phases of the open band signals are a
direct reflection of the initial conditions of the non-stationary
states.

It is also notable from Eqs. (43-44) that the open band (dichroic)
signal vanishes as we tune off-resonance\cite{foot6}, where the
imaginary lineshape function is negligible and ${\cal
C}_{k}(\omega )$ is vanishingly small. In contrast, the
off-resonant frequency dispersed signal is non-vanishing and
depends on the real part of the lineshape through ${\cal
S}_{k}(\omega )$. We also have $\varphi _{g}=\pi /2$ off-resonance
using the results of Appendix A. We then get
\begin{equation}
\widetilde{S}_{g}(\omega ,\tau )=\sum_{k=0}^{\infty }{\cal S}_{k}(\omega
)\sin \left( k\left( \omega _{0}\tau +\pi /2\right) \right) .  \label{67d}
\end{equation}
The dispersive functions $\Phi _{R}^{(k)}(\omega )$ are approximately
constant in the off-resonant limit. It then follows from Eq. (\ref{65b})
that the detuning dependence is mainly determined by the pulse
spectral function $\widetilde{G}_{p}$. The dispersed probe signal
consists of CARS and CSRS resonances that are centered near
$\omega \approx \omega _{c}\mp k\omega_{0}/2 $ and oppositely
phased\cite{walsh89,constantine97}.

The expressions derived here make minimal assumptions and can be
readily extended to incorporate optical heterodyne detection
schemes\cite{cho93a,cho93b,zeigler94,constantine97}. We have
derived explicit expressions for the dichroic response $\Im m
\{{\cal P}(t,\tau )\}$. It can be shown for the birefringent
response $\Re e \{{\cal P}(t,\tau )\}$, that the roles of the
quadrature amplitudes in Eq. (\ref{65x}) will be reversed; i.e.
that ${\cal C}_{k}(\omega )$ is the coefficient of the sine term
and ${\cal S}_{k}(\omega )$ is the coefficient of the cosine term.
Also, the CARS and CSRS contributions of ${\cal S}_{k}(\omega )$
will add constructively to give a non-vanishing birefringent open
band signal off-resonance. This has been observed in transparent
liquids\cite{constantine97}. For the resonant case, the results
derived here are a multilevel, high temperature generalization of
earlier treatments based on density matrix
pathways\cite{zeigler94} which considered a pair of vibrational
levels in the ground and excited states. In contrast to prior
treatments\cite{cho93a,cho93b}, no assumption has been made with
regard to the pulse durations in deriving Eqs. (38-42). Finally,
we note that although we have given expressions for only the
ground state response, analogous results are easily obtained for
the excited state response. Here, the emission lineshape function
$\Theta (\omega )$ in Eq. (\ref{41d}) plays a role similar to
$\Phi (\omega )$ in the ground state expressions.

\subsubsection{Reaction driven coherence}

As previously discussed, an advantage of the effective linear
response approach is that it allows for a rigorous calculation of
the probe response to non-radiatively driven coherence. As an
example, we consider the multilevel system depicted in Fig.\ 1(b),
and apply the effective linear response approach to the detection
of reaction induced coherent nuclear motion along the $Q$ degree
of freedom in the product state $\left| f\right>$.

The chemical reaction step and subsequent probe interaction
minimally constitute a three electronic level problem comprising
the electronic states $\left| e\right> $, $\left| f\right> $ and
$\left| f^{\prime }\right>$. The states  $\left|e\right>$ and
$\left| f\right> $ are non-radiatively coupled. The probe
interaction couples the ground state $\left| f\right> $ and
excited state $\left| f^{\prime }\right>$ of the product, and is
assumed to be well separated from the reaction step. The
Hamiltonian for the reaction-probe stage of the problem may be
written as
\begin{equation}
\widehat{H}_{NR}(t)=\pmatrix{ \widehat{H}_{f^\prime } &
-\mu_{ff^\prime }E_b(t) & 0 \cr -\mu_{f'f}E_b(t) & \widehat{H}_f
+U_f(R)  & J \cr 0 & J & \widehat{H}_e +U_e(R)}. \label{68a}
\end{equation}
Here, $U_{e,f}$ are the dissociative potentials along the
classical reaction coordinate $R$, for the states $\left|
e\right>$ and $\left| f\right>$. $\widehat{H}_{e,f}$ represent the
quantum degree of freedom ($Q$) coupled to the non-radiative
transition. $J$ is the non-radiative coupling parameter which, for
the case of MbNO photolysis discussed below, corresponds to the
spin-orbit coupling operator needed to account for the spin change
of the heme iron atom upon ligand photolysis. For simplicity, we
also assume that the pump interaction that couples the states
$\left| g\right>$ and $\left| e\right>$ is well separated from the
chemical reaction. The sequence of interactions and the relevant
coupling constants is then: pump, $\mu E_{a}\rightarrow $ chemical
reaction, $J\rightarrow $ and probe, $\mu E_{b}$. Each individual
interaction is assumed to occur independently. The separation of
these events allows us to consider the four electronic level
problem in Fig.\ 1(b) as three sequential two electronic level
problems.

We make the following two assumptions which are satisfied in the
case of MbNO photolysis: (i) Before the curve crossing takes
place, the system (in reactant state $\left| e\right>$) is assumed
be in thermal equilibrium along the $Q$ coordinate. (ii) The
quantum yield for the dissociative reaction along $R$ is assumed
to be unity; i.e. all the reactant molecules are transferred to
the product state during the surface crossing. Assumption (i)
holds if the $Q$ coordinate is not optically coupled to the
$\left|g\right>$ and $\left| e\right>$ electronic states. The pump
pulse merely transports a fraction of the ground state electronic
population to the excited state, leaving the vibrational state
along $Q$ unchanged. However, the $R$ degree of freedom is
dissociative on the excited state potential surface and is left
far from equilibrium after the pump excitation. For example, the
$220\ cm^{-1}$ Fe-His mode in MbNO is not optically coupled as
revealed by its absence in the resonance Raman spectrum of MbNO.
Its strong presence in the pump-probe signals demonstrates that
this mode is triggered into oscillation following the highly
efficient process\cite{schuresco78,gibson86,miller97} of ligand
photo-dissociation in MbNO\cite{ZHU94}.

A multidimensional Landau-Zener theory was previously developed to
describe vibrational coherence induced by non-radiative electronic
surface crossing\cite{zhu97} based on assumptions (i) and (ii)
above. For completeness of presentation, we have reproduced the
key results of that work in Appendix B. We also present
expressions for the amplitude $A_{f}$ and phase $\varphi _{f}$ of
the first moment of the nuclear motion induced on the product
surface $\left| f\right\rangle $. Using these reaction driven
initial conditions, we make the following representation for the
non-stationary nuclear density matrix in the product state
$\widehat{\rho }_{f}^{\prime }$;
\begin{equation}
\widehat{\rho }_{f}^{\prime }=
\widehat{D}(\lambda _{f})\widehat{\rho }_{T}^{(f)}
\widehat{D}^{\dagger }(\lambda _{f}),  \label{69a}
\end{equation}
where $\lambda _{f}=A_{f}e^{i\varphi _{f}}$ is the complex
displacement of the reaction induced coherence and $\widehat{\rho
}_{T}^{(f)}$ is the equilibrium thermal density matrix for the
nuclear Hamiltonian $\widehat{H}_{f}$. Since only the ground state
of the product $\left| f\right> $ is initially populated, the
probe response due to the nuclear dynamics in the product state
potential well can now be obtained along the same lines that led
to Eq. (\ref{50}). The correlation function $C_{f}$ for the
non-stationary response is obtained by replacing $A_{g},\
\varphi_g$ and $\Delta $ by $A_{f},\ \varphi_f$ and $\Delta_{f}$
in Eq. (\ref{50}). Here $\Delta_{f}$ is the dimensionless coupling
associated with the coupling of $Q$ to the $f\rightarrow f^{\prime
}$ transition. The equilibrium optical absorption correlator is
replaced by $K_{f}(t-t_{3})$, which refers to the pair of
electronic states $\left| f\right\rangle $ and $\left| f^{\prime
}\right\rangle $.

\section{Simulations and Discussion}

The numerical simulation of the open band signals involves
multiple integrals of the time correlation functions which can be
evaluated using standard algorithms. The full third order response
approach involves quadruple integrations over time. The effective
linear response approach replaces the correlation functions in Eq.
(\ref{27}) and (\ref{34}) by the analytic expressions in Eq.
(\ref{50}) and (\ref{54}). For field driven coherences, the
analytic expressions require the evaluation of the first moments
in Eqs. (\ref{41a}-\ref{41b}) and (\ref{47a}) of Appendix A. The
moments can be evaluated in a single step before doing the time
integrals. Hence, two of the time integrations are eliminated and
the overall computation time is significantly reduced. The signal
amplitude and phase in either of these approaches are obtained by
performing a Fourier transform of the delay dependent pump-probe
signal in Eq. (\ref{21}). An alternative method that can be used
when vibrational damping is neglected, is to directly calculate
the amplitude and phase of the dispersed and open band signals
using the analytic expressions derived in Sec. III.B.2. For
calculations involving only a few modes, the analytic formula for
the dispersed signal in Eq. (\ref {65}) can be easily evaluated,
and a subsequent integration over frequencies as in Eq. (\ref{10})
leads to the open band amplitude and phase.

In what follows, we demonstrate that amplitude and phase excitation profiles
of field induced coherence are equally well predicted by the third order
and the effective linear response approaches. We then apply the effective
linear response approach to calculate coherent signals induced by rapid
non-radiative reactions.

\subsection{\protect\smallskip Temperature and carrier frequency dependence}

One of the key parameters in a pump-probe experiment is the carrier
frequency of the pump and probe laser pulses. Not only is this parameter the
most easily accessible to the experimentalist, but measurements of the
oscillatory amplitude and phase profiles through the resonant region provide
crucial information regarding the origin of vibrational
coherence\cite{vos93,wang94,rosca00}. The computational simplicity of
the effective linear response approach, using the first moments, enables
a direct calculation of phase and amplitude profiles over the entire
absorption spectrum using realistic pulse widths.

In order to illustrate the accuracy and the physically intuitive aspects
of the effective linear response approach, we treat a simple
model system. Consider a single undamped mode with
$\omega _{0}=40\ cm^{-1}$ and $S=(\Delta^2/2)=0.5$ coupled to a
homogeneously broadened two level system with
$\Gamma _{e}=800\ cm ^{-1}$.
We assume $10\ fs$ pump and probe pulses in a degenerate
(same color for pump and probe) configuration.
The short pulse width allows comparison with the third order
response calculation which is quite formidable for long pulse
durations $\sim (50-100)\ fs$.

Expressing the delay dependent
oscillatory signal as
\begin{equation}
S(\tau )=A_{s}\cos (\omega _{0}\tau +\phi _{s}),  \label{xxx}
\end{equation}
we plot in Fig.\ 2 the degenerate open band amplitude $A_{s}$ and
the phase $\phi _{s}$ profiles for the $40\ cm^{-1}$ oscillations.
The ground and excited state profiles are plotted for a range of
carrier frequencies across the absorption maximum. The top panels
of Fig.\ 2(a) and (b) show the ground and excited state amplitude
and phase profiles for $T=0 \ K$ and $T=300 \ K$. Both the third
order response and the effective linear response outputs are
plotted. The excellent agreement between the two approaches is
evident over the entire range of carrier frequencies. We note that
the ground and excited state amplitudes dip near the classical
absorption and emission peaks, respectively, $\Omega _v$ and
$\Omega _v-2\omega _{0}S$. The phase of the ground state signal
shows a variation of $2\pi $ as the carrier frequency is detuned
across $\Omega _{v}$. There is a sharp jump between $\pm (\pi /2)$
at $\Omega _{v}$. In contrast to the ground state, the excited
state phase remains constant apart from a $\pi $ phase jump at
$\Omega _{v}-2\omega _{0}S$. At low temperature, the amplitude of
the ground state signal drops by almost an order of magnitude. The
approach of the phase toward $\pm (\pi /2)$ on either side of the
discontinuity at $\Omega _{v}$ is steady and almost linear. For
the high temperature case, the approach of the phase toward $\pm
(\pi /2)$ occurs much more sharply near $\Omega _{v}$. While the
excited state phase is independent of temperature, the excited
state amplitude becomes more asymmetric as the temperature is
increased, as seen in Fig.\ 2(b).

All of the above aspects of the ground and excited state signals
can be clearly understood using the first moments of the pump
induced oscillations; see Appendix A. First, recall that the open
band phase yields the initial phase of the pump induced
oscillation to within an additive factor of $\pi $ as in Eq.
({\ref{68cc}). This is shown for the ground state signal in Fig.\
3. We plot the open band phase of the $40\ cm^{-1}$ mode over an
expanded range that includes the off-resonant limit, along with
the initial phase of the wave-packet calculated using Eqs.
(\ref{41a}) and (\ref{41b}).
\begin{figure}[htbp]
\begin{center}
\mbox{\epsfig{file=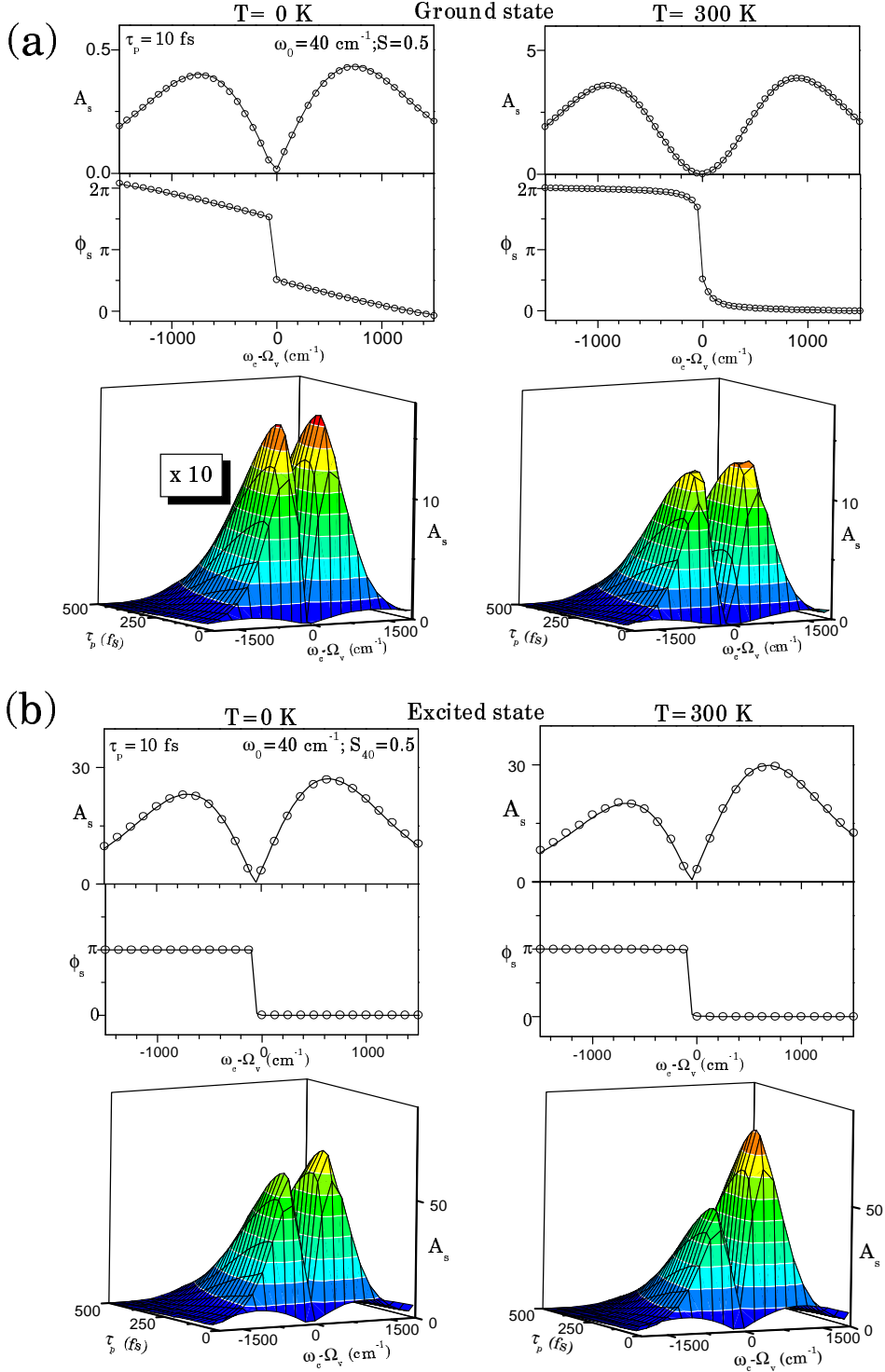,height=200mm}} \caption{Comparison
of one-color pump-probe open band amplitude $A_{s}$ and phase
$\phi _{s}$ profiles predicted by the first moment based effective
linear response approach (solid line) and the third order response
approach (circles). The profiles shown are for a single undamped
mode ($\omega _{0}=40 \ cm^{-1},\ S=0.5$) coupled to a
homogeneously broadened ($\Gamma _{e}=800 \ cm^{-1}$) two level
system at $T=300 \ K$ and $T=0 \ K$. The top panels of (a) and (b)
show the amplitude and phase profiles for pulse width of $10 \
fs$. The bottom panels of (a) and (b) show the amplitude of the
ground and excited state signals as two dimensional functions of
the pulse width $\tau _{p}$ and the pulse carrier frequency
$\omega _{c}$, for both $T=0 \ K$ and $T=300 \ K$. The analytic
expressions in Eq. (38-42) were employed in the 2-D simulations.}
\end{center}
\end{figure}
Fig. 3(c) schematically depicts the effective initial conditions
prepared by the pump pulse in the ground and excited states. Shown
are the initial conditions for three different pump carrier
frequencies, near resonance (indicated by downward pointing arrows
in Fig.\ 3(a)) and off-resonance towards the red side of the
absorption maximum. The arrows above the wave-packets indicate the
direction of the momentum induced by the pump pulse. From Eqs.
(\ref{41a}-\ref{41b}), we note that the mean nuclear position and
momentum in the ground state depend on the derivative of the
absorption and dispersion lineshapes respectively\cite{kumar00}.
Thus, the wave-packet is undisplaced from equilibrium for
excitation at $\omega _{c}=\Omega _{v}$. The momentum attains a
maximum value at this frequency, and is signed opposite to the
excited state equilibrium position
shift\cite{kumar00,smith94,jonas95}.

\begin{figure}[htbp]
\begin{center}
\mbox{\epsfig{file=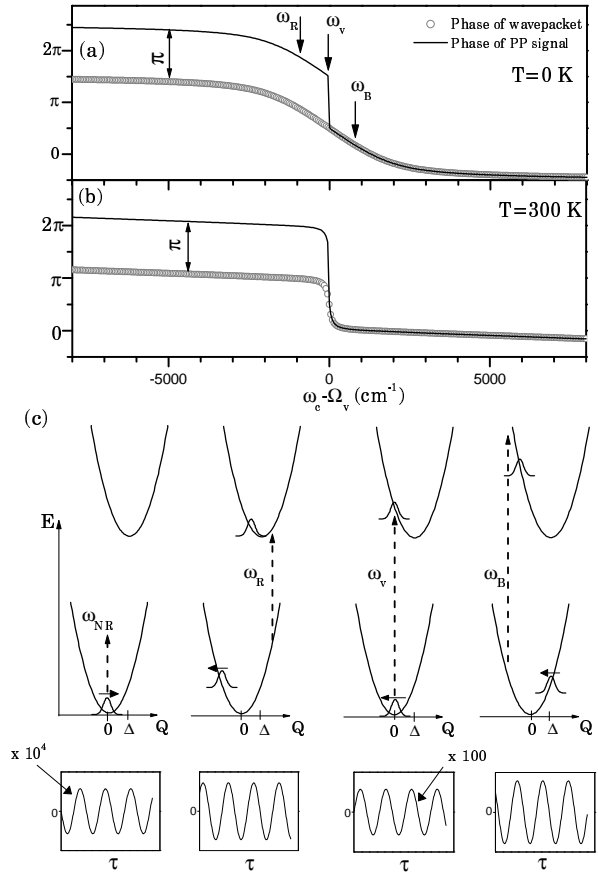,height=120mm}} \caption{Interpretation
of the open band phase for the ground state signal in terms of the
first moments of the pump induced nuclear wavepackets. (a) The
phase $\phi_s$ of the 40$ \ cm^{-1}$ mode (from the example in
Fig.\ 2) at $T=0 \ K$ is plotted (solid line) along with the
initial phase $\varphi_g$ of the pump induced wavepacket
(circles). (b) Same as in (a) but for $T=300 \ K$. (c) Schematic
of the pump induced effective initial conditions in the ground and
excited states in a two level system, shown for three different
pump/probe carrier frequencies across the absorption maximum
$\Omega _{v}$, and for non-resonant excitation $\omega _{NR}$. The
arrows above the wavepackets indicate the direction of the
momentum imparted. The panels directly below the potential curves
show the corresponding oscillatory signal as a function of
pump-probe delay time.}
\end{center}
\end{figure}
As depicted in Fig. 3, when the pump pulse carrier frequency is
tuned toward the red side of the absorption maximum $\omega _{R}$,
the ground state wave-packet is displaced toward decreased
bond-lengths and receives a momentum kick in the same direction.
Thus, a probe pulse of carrier frequency $\omega_{R}$ incident on
the sample sees an initially depleted nuclear distribution
(bleach) in the ground state. The probe difference transmission
signal (pump-on minus pump-off) will at $\tau =0$ be positive and
will simultaneously have a positive slope due to the sign of the
wave-packet momentum. This is seen from the delay dependent signal
plotted directly below the potentials in Fig.\ 3(c). The phase of
the oscillatory signal $S(\tau )$ in Eq. (\ref{xxx}) will obey
$(3\pi /2)<\phi _{s}<2\pi $. On the other hand, for blue detuning
from absorption maximum $\omega _{B}$, the initial transmission
signal will again be positive due to a depleted nuclear
distribution in the probed region. However, the signal has a
negative slope at $\tau =0$, since now the momentum of the
wave-packet points toward the probing region. In this case the
signal phase will obey $0<\phi _{s}<(\pi/2)$ and is also precisely
in phase with the wave-packet itself. When the carrier frequency
is tuned to resonance at $\omega _{v}$, the pump pulse merely
imparts a momentum to the wave-packet so that the wave-packet
phase is precisely $(\pi /2)$. As described below, the CSRS and
CARS resonances are oppositely phased for excitation at resonant
maximum. The integrated signal therefore vanishes at $\omega_v$
and the phase becomes undefined. Incrementally to the red and blue
sides of the absorption maximum, the open band phase shows a
discontinuous jump of $\pi $ since the momentum impulse always
points in the same direction in the resonance region. Note that
Fig.\ 3(c) shows, for $\omega_{v}$, the oscillatory signal
incrementally to the red of the absorption maximum, before the
phase jump.

A similar analysis is possible for the excited state signal. The
excited state response arises from the stimulated emission (by the
probe) from the vibrationally coherent excited state. Since the
excited state wave-packet oscillates about the shifted equilibrium
position $\Delta $, the amplitude dip and phase flip of the
pump-probe signal from the excited state occurs near the classical
emission peak at $\Omega_{v}-2\omega _{0}S$. As depicted in Fig.\
3(c), the excited state wave-packet does not receive a momentum
impulse; see Appendix A. Furthermore, it is created on the same
side of the excited state harmonic well irrespective of the pump
carrier frequency\cite{kumar00}. Thus, the phase of the excited
state wave-packet is independent of the pump carrier frequency.
The corresponding signal phase is fixed at $0$ for blue detuning
where there is increased stimulated emission at $\tau =0$. The
phase is fixed at $\pi $ for red detuning where the stimulated
emission is minimum at $\tau =0$. Note in Fig.\ 2(b) the nearly
order of magnitude increase in the amplitude of the excited state
signal compared to the ground state. The increase can be traced to
the fact that for impulsive excitation using weak fields, the
excited state wave-packet is situated very close to the vertical
energy gap $\Omega _{v}$. This corresponds to the ground state
hole that has a displacement much smaller than
$\Delta$\cite{banin94,smith96,kumar00}. Thus, the amplitude of
excited state oscillations about the equilibrium position $\Delta
$ is much larger than that of ground state oscillations about
zero. The excited state coherence will thereby be the dominant
contribution for systems with long-lived excited
states\cite{tang86,zewail88,Fragnito89,scherer91,vos93,yang99}.
The asymmetry in the excited state amplitude profile can be seen
in Fig.\ 3(c) to arise from the fact that the wave-packet
amplitude in the excited state is larger for blue excitation than
for red\cite{kumar00}.

The temperature dependence of the oscillatory amplitude and phase
in Fig.\ 2 can be understood via the relative temperature
dependence of the initial position and momentum induced by the
pump pulse. For high temperatures, the mean thermal phonon
population $\overline{n}>>1$ and $Q_{g0}$ is enhanced according to
 Eq. (\ref{41a}), so that the complex displacement $\lambda
_{g}=\left( Q_{g0}+iP_{g0}\right) /\sqrt{2}$ is dominated by the
real part. Hence the initial phase of the wavepacket stays closer
to zero or $\pi$ at higher temperatures, except near $\Omega
_{v}$, where $Q_{g0}$ vanishes. Correspondingly, the open band
phase also stays closer to zero (or $2\pi $) except for the
discontinuous transition near $\Omega _{v}$. When the temperature
is lowered, $Q_{g0}$ drops sharply since $\overline{n}<<1$ and
becomes comparable in magnitude to $P_{g0}$. Thus, at low
temperatures, the amplitude of the ground state signal drops
dramatically, and the phase varies almost linearly on either side
of the discontinuity at $\Omega _{v}$ as in Fig.\ 2(a). The
temperature dependence of the excited state first moment $Q_{e0}$
is primarily determined by the relative magnitude of $\Phi
_{I}(\omega )$ and the difference lineshape $\widehat{\Delta }\Phi
_{I}(\omega )$, which appears as the coefficient of $\overline{n}$
in Eq. (\ref{47a}). The difference lineshape is responsible for
enhancing the asymmetry in the amplitude profile at high
temperatures as in Fig.\ 2(b).

An important observation to be made from Fig.\ 3(c) is that for
the linearly displaced oscillator model, the amplitude and phase
profiles of field driven coherence are independent of the sign of
the electron nuclear coupling $\Delta $. If $\Delta $ were
negative, then the same arguments given above would apply. The
amplitude and phase behavior would be the same as in Fig.\ 2. This
can be seen directly from the fact that the non-linear response
functions in Eqs. (\ref{28}-\ref{29}) and (\ref{35}-\ref{36})
depend only on $\Delta ^{2}$ through the function $g(s)$. However,
the schematic depicted in Fig.\ 3 clearly indicates that the
absolute sign of the potential displacements are not revealed in a
pump-probe experiment.

The depiction of pump induced initial conditions in Fig.\ 3
contradicts prior predictions arising from time dependent
wave-packet pictures of impulsive stimulated light
scattering\cite{pollard90,pollard92,dexheimer00}. The prior work
suggests that the ground state wave-packet is always created on
the side of the ground state well that is closer to the excited
potential minimum. This is {\em not} correct, as is shown by a
careful analysis of the pump pulse interaction
\cite{kumar00,shen99,smith94}. It is clear from Fig.\ 3(c) that
the centroid of the ground state wave-packet induced by impulsive
excitation is strongly sensitive to the carrier frequency of the
laser pulse.

From Figs. 2 and 3 we conclude that the moment analysis based
approach offers a physically intuitive and accurate interpretation
of the observed amplitude and phase behavior of open band FCS
signals. Another significant advantage offered by this approach is
that the analytic form of the effective linear response functions
reduces the computation times. For the comparisons made in Figs.\
2 and 3 (top panels), we chose a very short $10 \ fs$ pulse in
order to make the calculations using the third order approach
possible with reasonable computation times. However, in many
experimental situations, including those reported here, the pulse
width is often much longer ($\sim 100 \ fs)$. For long pulse
widths the two extra integrations involved in the third order
approach makes the computation quite formidable. In Fig.\ 4, we
compare the computation times for the linear response and third
order response calculations as a function of pulse width $\tau
_{p}$. The advantage offered by the effective linear response
approach is clear, especially when it is of interest to calculate
the amplitude and phase profiles at many carrier frequencies. The
computation time for calculating the open band amplitude and phase
using the analytic formulae in Eqs. (38-42) for a single mode and
a two mode case is also plotted in the figure. The integral of the
analytic expressions is over the spectral profiles of the probe
pulse so that a larger value of $\tau _{p}$ implies a narrower
range of integration. In contrast to the numerical integrations in
time domain, the computation time in this case therefore decreases
with $\tau _{p}$. As the number of modes involved increases, the
summation in Eq. (\ref{65}) involves multiple indices and the
calculation becomes more complex. However, the usefulness of the
analytic expression for a few modes is evident from Fig.\ 4.
\begin{figure}[htbp]
\begin{center}
\mbox{\epsfig{file=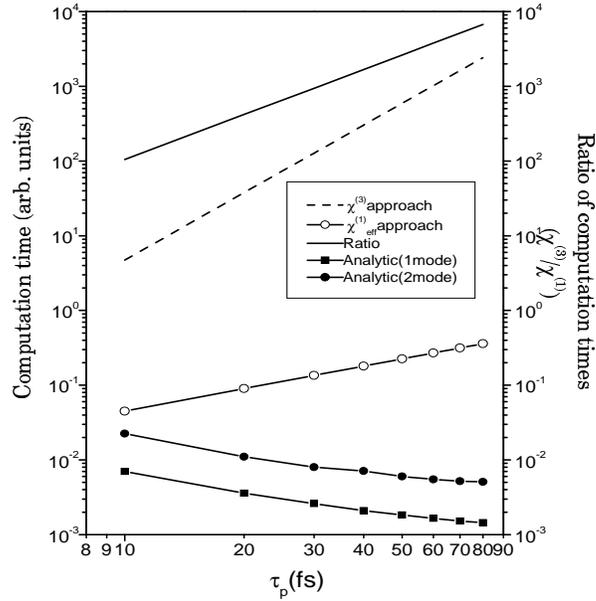,height=100mm,width=80mm}}
\caption{Comparison of computation times for the calculation of
the one-color ground state pump-probe signal, using the third
order response approach and the effective linear response
approach, as a function of pulse width. Both the absolute time to
calculate the signal for 40 $\omega _{c}$ points and the ratio of
the computation times for the two approaches are plotted. Also
shown is the computation time for calculating the profiles using
the analytic expressions for the pump-probe signal in Eqs.
(38-42). }
\end{center}
\end{figure}
As an illustration of the use of the analytic expressions, we
calculate the ground and excited state amplitude profiles for a
wider range of pulse widths $(5-500) \ fs$ in the bottom panels of
Fig.\ 2(a) and (b). The three-dimensional plots are useful in
capturing the behavior of pump-probe signals over a wide range of
pulse widths and carrier frequencies. It is seen that the
temperature dependence of the ground state signal amplitude is
rather uniform over the entire manifold of $\tau _{p}$ and $\omega
_{c}$. The excited state signal shows more interesting behavior,
with the asymmetry in the amplitude profiles more pronounced at
longer pulse widths. Both the ground and excited state amplitudes
vanish in the limit of very short and very long (compared to the
vibrational period) pulses, attaining an optimum for some
intermediate value of  $\tau _{p}$. The vanishing of the ground
state signal for $\tau _{p} \rightarrow 0$ can be traced to the
fact that the ground coherence vanishes in the short pulse
limit\cite{jonas95,pollard92,pollard92b,kumar00}. It is
interesting to note from Eq. (\ref{47a}) that the excited state
nuclei are however coherent for $\tau _{p} \rightarrow 0$. It can
be shown that the corresponding density matrix is simply thermal
density matrix $\widehat{\rho}_{T}^{(e)}$ placed vertically above
$Q=0$ on the excited state potential well\cite{kumar00}. Note that
the representation in Eq. (\ref{4}) is essentially exact in this
limit. The vanishing of the excited state open band signal for
$\tau _{p} \rightarrow 0$ occurs rather due to the fact that the
open band signal for a spectrally broad probe is essentially the
integral over the derivative-like basis functions, $\Theta
_{I,R}^{(k)}(\omega )$, which are analogous to $\Phi
_{I,R}^{(k)}(\omega )$ plotted in Fig.\ 5 (see below).

Finally, we consider a two color pump-probe experiment, which
employs a fixed carrier frequency for the pump and a variable
carrier frequency of the probe. The analysis presented so far
applies equally well to two color pump-probe experiments. In this
case, the pump imparts a fixed initial condition to the
wave-packet that is determined from the results of Appendix-A. The
signal amplitude and phase profiles as a function of the probe
carrier frequency will be similar to that of the excited state
degenerate pump-probe signal. The amplitude will exhibit a dip at
$\Omega _{v}$, and the signal phase will be the initial
wave-packet phase, but for a $\pi $ jump at $\Omega _{v}$. Since
the pump induced initial conditions are independent of the probe
carrier frequency $\omega_c$, the computation of two color
pump-probe profiles is less cumbersome than the degenerate
pump-probe profiles.

\subsection{Dispersed pump-probe measurements and off-resonant response}

When the pump pulse is tuned off-resonance,
$\widehat{\Delta}\Phi_{I}(\omega )\rightarrow 0$ and Eqs.
(\ref{41a}-\ref{41b}) show that the ground state wave-packet
merely receives a momentum kick, and is undisplaced from its
equilibrium position. Thus, the phase of the open band signal
approaches $\pm (\pi /2)$ as seen from Figs.\ 3(a) and (b).
Furthermore, the dependence of the wave-packet momentum on
$\widehat{\Delta }\Phi _{R}$ implies that the momentum impulse
changes direction as we begin to move off-resonance on the red and
blue sides of the absorption maximum. Correspondingly, the signal
phase crosses $2\pi$ (zero) into the first (fourth) quadrant as
seen in Fig.\ 3(a) and (b). The approach of the open band phase
toward the off-resonant limit of $\pm (\pi /2)$ is more dramatic
at lower temperatures, where $P_{g0}$ is comparable in magnitude
to $Q_{g0}$. At high temperatures, the magnitude of $Q_{g0}$ is
dramatically enhanced relative that of $Q_{g0}$ because of the
thermal factor in Eq. (\ref{41a}). This enhances the contribution
of $\widehat{\Delta}\Phi_{I}(\omega )$ to the signal phase. The
off-resonant limit is thus attained at further detuning from
absorption maximum at high temperatures.

While the open band (dichroic) signal approaches zero
off-resonance, we showed earlier that the dispersed probe signal
does not vanish in this limit and is related to $\Phi _{R}(\omega
)$ as in Eq. (\ref{67d}). In Fig.\ 5, we demonstrate the on- and
off-resonance behavior of the dispersed signal, using the same
model parameters as in Fig.\ 2. In the top panel, we plot the
basis functions for the $T=0$ ground state fundamental $(k=1)$
oscillations $\Phi _{I}^{(1)}(\omega )$ and $\Phi_{R}^{(1)}(\omega
)$ given by Eqs. (\ref{66}-\ref{67}). The basis functions for
$k=1$ are seen to behave roughly as the first derivatives of the
absorption and dispersion lineshapes. Similarly, the basis
functions for the first overtone $\Phi _{I}^{(2)}(\omega )$ and
$\Phi _{R}^{(2)}(\omega )$ can be shown to behave like the second
derivatives of the lineshapes and so on. The Gaussian spectral
profile of the laser pulse ($\tau_p=50\ fs$) used in the
calculation is also shown in the upper panel for three different
carrier frequencies $\omega _{c1}, $ $\omega _{c2}$, and $\omega
_{c3}$. The functions ${\cal C}_{1}(\omega )$ and ${\cal
S}_{1}(\omega )$ and the dispersed probe amplitude and phase
$\widetilde{A}_{1}(\omega )$ and $\widetilde{\phi }_{1}(\omega )$
are plotted directly below. As specified in Eqs. (\ref{65a}) and
(\ref{65b}), the quadrature amplitudes ${\cal C}_{1}(\omega )$ and
${\cal S}_{1}(\omega )$ are obtained by superposing the Stokes and
anti-Stokes shifted product functions $G_{p}(\omega _{d},\omega
_{0})\Phi_{I,R}^{(1)}(\omega )$ and $G_{p}(\omega _{d},-\omega
_{0})\Phi_{I,R}^{(-1)}(\omega )$, which correspond to CSRS and
CARS resonances respectively. While the (dichroic) Stokes and
anti-Stokes components are additive for ${\cal C}_{1}(\omega )$
they are subtractive for ${\cal S}_{1}(\omega )$. It is clear from
Fig.\ 5 that at the absorption maximum $\omega _{c}=\omega _{c1}$,
the derivative nature of $\Phi _{I}^{(1)}(\omega)$ gives rise to
Stokes and anti-Stokes components of ${\cal C}_{1}(\omega )$ that
are oppositely signed on either side of the carrier frequency.
Thus, the dispersed signal is $\pi $ out of phase for red and blue
detuning from the carrier frequency. The integral of both ${\cal
C}_{1}(\omega )$ and ${\cal S}_{1}(\omega )$ vanishes, giving rise
to a dip in the open band signal amplitude. When the carrier
frequency is tuned near the shoulder of the absorption spectrum
$\omega _{c}=\omega _{c2}$, the Stokes and anti-Stokes components
of ${\cal C}_{1}(\omega )$ add constructively to give a single
peak. ${\cal S}_{1}(\omega )$ is very small compared to ${\cal
C}_{1}(\omega )$ since $\Phi _{R}^{(\pm 1)}(\omega )$ are small in
this region. For off-resonant excitation $\omega _{c}=\omega
_{c3}$, the dispersive term ${\cal S}_{1}(\omega )$ begins to
dominate and the signal approaches the limit in Eq. (\ref{67d}).
While the details of the behavior of the dispersed probe signal
are highly mode specific\cite{zhou99}, the simulations shown in
Fig.\ 5 illustrate the general aspects of the dispersed pump-probe
signals based on equilibrium lineshape functions.

It is clear from Fig.\ 5 that the direct connection with
equilibrium lineshapes made in Eqs. (38-42) allows a clear picture
of the dispersed probe measurements. Furthermore, when dealing
with complex multimode systems, one can directly use the
experimentally measured absorption lineshape and its Kramer-Kronig
transform, the dispersive lineshape. The need for modeling the
detailed mechanisms responsible for line broadening is thus
avoided. However, additional steps are required when inhomogeneous
broadening is significant (see below). Generally, this approach is
analogous\cite{foot3} to the use of transform methods in the
calculation of resonance Raman scattering
cross-sections\cite{page81,Stallard83,Schomacker89,page91}.

\begin{figure}[htbp]
\begin{center}
\mbox{\epsfig{file=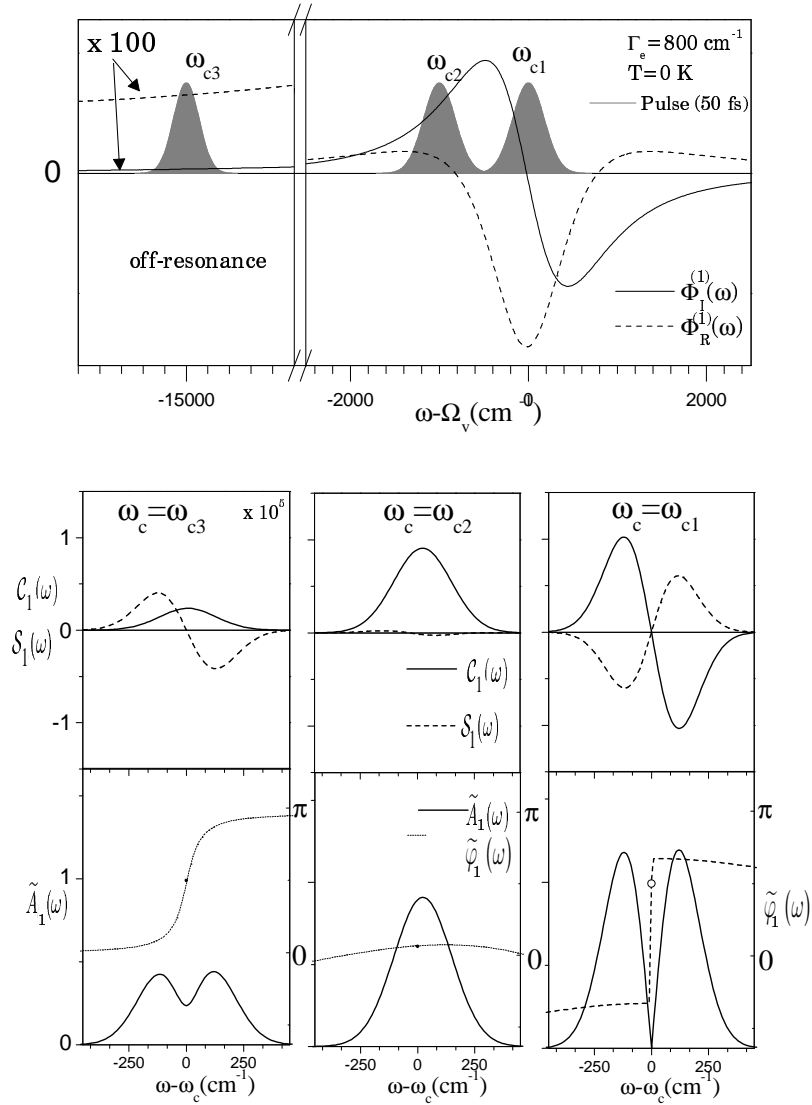,height=150mm}} \caption{Dispersed
pump-probe signal for the two electronic level system considered
in Fig.\ 2. The top panel shows the absorptive and dispersive
basis functions $\Phi^{(1)}_{I}(\omega)$ and
$\Phi^{(1)}_{R}(\omega )$ defined in Eqs. (\ref{66}) and
(\ref{67}), for $T=0 \ K$. The spectral profile of the pulse
($\tau_p=50 \ fs$) is shown for three different pulse carrier
frequencies. The bottom panels show the corresponding quadrature
amplitudes ${\cal C}_{1}(\omega )$ and ${\cal S}_{1}(\omega )$,
and the amplitude $\widetilde{A}_{1}(\omega )$ and phase
$\widetilde{\phi }_{1}(\omega )$ of the $40 \ cm^{-1}$ fundamental
of the dispersed signal. The open band phase for each carrier
frequency is also shown (filled circle).}
\end{center}
\end{figure}
\subsection{Effects of inhomogeneous broadening}

Until now, we have assumed that the molecular system under
consideration has a fixed electronic energy gap ($\Omega _{00})$
and a homogeneously broadened absorption spectrum. The homogeneous
broadening mechanism reflects the electronic population decay as
well as the rapid fluctuations that take place between the system
and the environment. In addition to the homogeneous contribution,
the width of the absorption lineshape can also have a quasi-static
origin. For example, a distribution of electronic transition
frequencies arising from local changes in the environment or
disorder in the system\cite {srajer86}. Several spectroscopic
techniques exist\cite {moerner93,becker89,personov}, which can be
used to selectively probe the effects of inhomogeneous broadening.
In this section, we consider the effects of inhomogeneous
broadening on pump-probe signals. Experimental examples of open
band phase measurements on myoglobin (Mb) are presented to
demonstrate the presence of an asymmetric inhomogeneously
broadened absorption spectrum for Mb.

In the presence of inhomogeneous broadening, the homogeneous
pump-probe signal must be averaged over the inhomogeneous
distribution ${\cal F}_{I}(\Omega _{00})$. In the third order
response approach, this averaging is simply a Fourier transform of
the function ${\cal F}_{I}$ and can be carried out independently
of the time integrations over the pump fields\cite{muk95}; see Eqs
(\ref{x3}) and (\ref{x4}) of Appendix-C. In the effective linear
response approach, however, the averaging over ${\cal
F}_{I}(\Omega _{00})$ cannot be carried out as in Eqs. (\ref{x3})
and (\ref{x4}). The time integrations over $t_{1}$ and $t_{2}$
involved in the pump interaction are carried out separately to
find the first moments. $\Omega _{00}$ appears through the
lineshape functions $\Phi _{I}(\omega )$ and $\Phi _{R}(\omega )$
in the first moment expressions, Eqs. (\ref{41a}-\ref{41b})). In
order to include the inhomogeneities, we first calculate the
homogeneous pump-probe signal and then numerically average the
signal over the distribution ${\cal F}_{I}(\Omega _{00})$. When
experimentally measured lineshapes are used in place of the time
correlator expressions, the homogeneous lineshape must be
deconvolved from the measured lineshapes before being used in the
pump and probe steps of the calculation. The inhomogeneous
pump-probe signal is obtained by a final convolution of the
homogeneous pump-probe signal with the inhomogeneous distribution.

Inhomogeneous broadening affects the ground and excited state
responses in a distinct fashion, owing to their contrasting phase
behavior. Recall that the oscillatory phase of the ground state
signal is approximately in phase on either side of the resonant
maximum $\Omega _{v}$. Since the inhomogeneous pump-probe signal
is essentially a superposition of homogeneous signals, the
inhomogeneous amplitude will be non-vanishing at $\Omega _{v}$.
The phase discontinuity will also be smeared out in the presence
of inhomogeneity.
\begin{figure}[htbp]
\begin{center}
\mbox{\epsfig{file=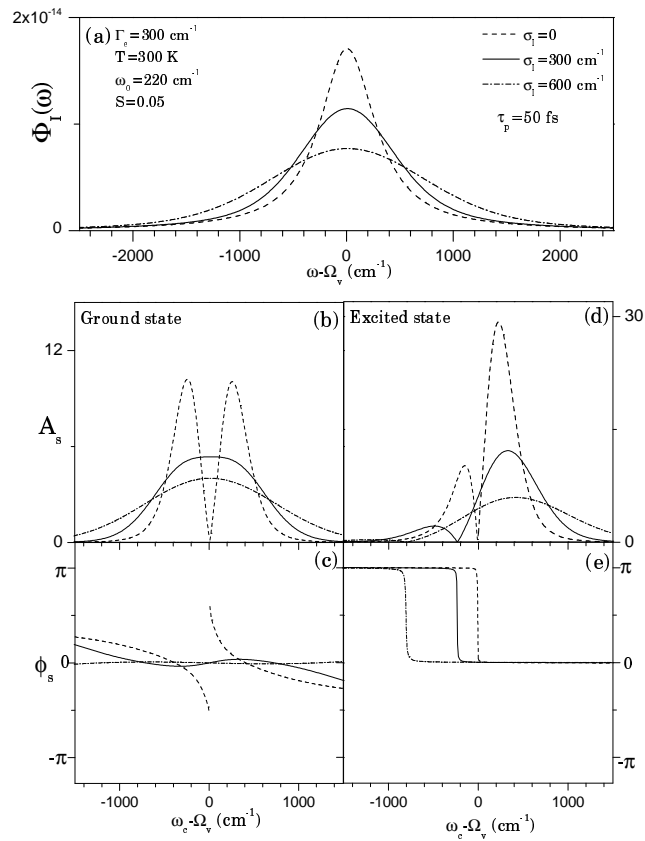,height=120mm}} \caption{Effects of
inhomogeneous brodening on one color pump-probe amplitude and
phase profiles. (a) shows the absorption lineshape function for
the homogeneous case (dashed line), and with Gaussian
inhomogeneities of $\sigma _{I}=300 \ cm^{-1}$ (solid line) and
$\sigma_{I}=600 \ cm^{-1}$ (dash dot line). The amplitude and
phase profiles for the ground state case are shown in panels (b)
and (c), for the three lineshapes. The corresponding excited state
amplitude and phase are shown in panels (d) and (e). }
\end{center}
\end{figure}
In Fig.\ 6, we consider a homogeneous lineshape
with $\Gamma _{e}=300\ cm^{-1}$ and a single undamped $220\
cm^{-1}$ mode with coupling $S=0.05$. Pump-probe signals are
calculated for $T=300 \ K$ using both the homogeneous lineshape
and with Gaussian inhomogeneities included. The top panel in the
Fig.\ 6 shows $\Phi_I(\omega)$ with three values of the
inhomogeneous width $\sigma_I=0,\ 300\ cm^{-1},\ {\rm and} \ 600\
cm^{-1}$. The lower panels show the amplitude and phase of ground
and excited state signals for the three cases assuming a pulse
width of $\tau_{p}=50\ fs$.

It is seen that while the amplitude dip and phase discontinuity of
the ground state signal are dramatically removed by the
inhomogeneity, they exhibit a red-shift for the excited state. The
shift of the excited state signal arises from the asymmetry of the
excited state amplitude: since the amplitude of the homogeneous
stimulated emission signal on the red side is much smaller than on
the blue side of $\Omega _{v}$, an effective cancellation of the
positive and negative lobes of the signed amplitude will occur
towards the red.
\begin{figure}[htbp]
\begin{center}
\mbox{\epsfig{file=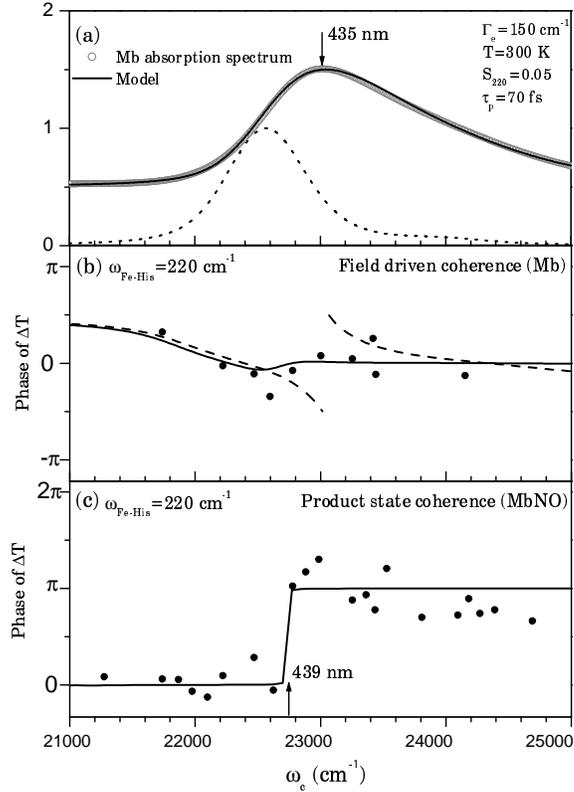,height=120mm}} \caption{(a)
Equilibrium absorption spectrum of Mb at $T=300 \ K$ (open circle)
shown with the theoretical fit (solid line) based on a prior model
for the asymmetric deoxy Mb lineshape. The underlying homogeneous
lineshape (before convolution with inhomogeneities) is also shown
(dotted line). (b) Experimental Soret band phase profile for $220
\ cm^{-1}$ mode coherence from deoxy Mb (filled circles) is shown
with the theoretical fit based on an inhomogeneous model (solid
line) and a homogeneous model (dashed line) using the asymmetric
lineshape plotted in (a). (c) Experimental phase profile for the
$220 \ cm^{-1}$ mode in MbNO for a range of carrier frequencies
across the Soret band is shown with a theoretical fit based on
simple model for reaction driven coherence in MbNO.}
\end{center}
\end{figure}

\subsubsection{Example: Inhomogeneous broadening in myoglobin}

The characteristic features of the ground and excited state
amplitude and phase profiles can be used as indications of
inhomogeneous broadening. As an example, consider the heme protein
deoxy myoglobin (Mb), which is an oxygen storage protein found in
muscle cells. Mb possesses a highly asymmetric and broad
absorption spectrum (Soret band) in its ligand-free, high spin
(deoxy, S=2) state. In previous studies, the Soret band of Mb was
modeled using a non-Gaussian distribution of electronic energy
levels, ascribed to disorder in the position of the central iron
atom of the porphyrin ring\cite{srajer86,champ92}. In pump-probe
and resonance Raman experiments on Mb, the Soret excited state is
found\cite{champ80,rosca00} to be very short lived ($\leq 30 \
fs$). Thus, the detected pump-probe signals for deoxy Mb are
presumably due to ground state coherence driven by impulsive
stimulated Raman scattering. Here, we focus attention on the
Fe-His mode at $220 \ cm^{-1}$ which is a prominent mode in the
resonance Raman experiments on Mb.

In the top panel of Fig.\ 7, we plot the experimental absorption
spectrum of Mb along with a theoretical fit based on the
previously proposed model for the inhomogeneous
distribution\cite{srajer86,champ92}. All resonance Raman active
modes are included in the model, with coupling strengths
determined from their absolute resonance Raman
cross-sections\cite{srajer91}. The pump-probe calculations are
carried out using the effective linear response approach using a
realistic pulse width of $70 \ fs$. For simplicity, only the $220
\ cm^{-1}$ mode $(S=0.05)$ was taken to be displaced from
equilibrium. In Fig.\ 7(b), we plot the phase profile of the $220
\  cm^{-1}$ mode observed in open band (one color) pump-probe
measurements on Mb across the absorption maximum. The theoretical
prediction for the phase of the $220 \ cm^{-1}$ mode in the ground
state is shown for both the inhomogeneous lineshape
model\cite{srajer86,champ92} as well as for a homogeneous model
using the observed lineshape.

It is clear from Fig.\ 7(b) that the theoretical prediction using
the inhomogeneous lineshape is in good agreement with the
experimental data. Note that the phase approaches the off-resonant
limit of $(\pi /2)$ much more rapidly on the red side than on the
blue side of the absorption maximum. This reflects the strong
asymmetry in the absorption spectrum. The sharp phase jump between
$-(\pi /2)$ to $+(\pi /2)$ predicted by the homogeneous model is
clearly not observed in the experimental data. Also, The amplitude
of the 220$\ cm^{-1}$ mode does not show a clear dip near
absorption maximum, contrary to the prediction of the homogeneous
model. These observations strongly support the presence of
significant inhomogeneous broadening in deoxy Mb, as well as the
ground state origin of the vibrational oscillations. If the $220\
cm^{-1}$ coherence originated from the excited state rather than
the ground state, an amplitude dip and phase flip should be
observed but red-shifted from the emission maximum as discussed
above.

\subsection{Reaction driven coherence:\\{\it Coherent oscillations in MbNO}}

We now apply the effective linear response approach to coherence
driven by a chemical reaction of the type shown in Fig. 1(b). The
effective linear response function for a two level system
consisting of the product ground and excited states $\left|
f\right\rangle $ and $\left| f^{\prime}\right\rangle $ was
discussed in section III.B.3. The nuclear density matrix for the
state $\left|f\right\rangle$ is described by a displaced thermal
density matrix. The displacements are the first moments of the
vibrational coherence driven by a Landau-Zener surface crossing
between the non-radiative states $\left| e\right\rangle $ and
$\left| f\right\rangle $ presented in Appendix-B. The pump-probe
signal arises from the vibrationally coherent state $\left|
f\right\rangle $. As discussed earlier, we are concerned with the
situation where the vibrational mode $Q$ is not coupled to the
$g\rightarrow e$ optical excitation, but is triggered solely by
the non-radiative chemical reaction between $\left| e\right>$ and
$\left|f\right>$ that follows pump excitation. For example, $Q$
could represent the 220$ \ cm^{-1}$ Fe-His mode in the reactive
sample MbNO. The 220$\ cm^{-1}$ mode is not resonance Raman active
in the reactant ground state ($\left|g\right\rangle $) but is
nevertheless observed in pump-probe photolysis signals.

The spectral lineshape function also plays an important role in
the multidimensional Landau-Zener theory\cite{zhu97}. As specified
in Eq. (\ref{b3}) of Appendix-B, the transition rate between the
states $\left|e\right\rangle $ and $\left| f\right\rangle $ is
obtained by taking the zero frequency limit of the spectral
function for the $e\rightarrow f$ transition i.e., $\Phi
_{I}^{ef}(\omega =0,\Omega _{00}^{ef}(R(s)))$. Here, $\Omega
_{00}^{ef}(R(s))$ is the time dependent energy gap between the
$\left| e\right\rangle $ and $\left| f\right\rangle $ states as
shown in Fig.\ 1(b) and $R(s)$ is the time dependent
semi-classical reaction coordinate. As the reaction coordinate
sweeps through the crossing point, $\Omega_{00}^{ef}(R(s))$ passes
through zero. The reaction rate ${\cal G}(R(s))$ is integrated as
in Eq. (\ref{b2}) to obtain the time dependent quantum yield
$P_{ef}(t)$. In the present discussion, we hold to the assumption
of constant reaction velocity at the crossing point. This means
that the energy gap $\Omega _{00}^{ef}(R(s))$ is decreasing at a
constant rate $\beta $.

Our aim here is to illustrate the analysis of pump-probe signals
associated with the reaction dynamics, and we restrict our
attention to the role played by the reaction velocity $\beta $. We
refer the reader to reference\cite{zhu97} for the details of the
role played by the various reaction specific parameters in the
population and coherence dynamics. In Fig.\ 8(a), we simulate the
spectral function $\Phi _{I}^{ef}$ for the non-radiative states.
The coupling of the $220 \ cm^{-1}$ mode is taken as $S=0.5$
corresponding to the $0.1\AA $ Fe-His bond length contraction
associated with the loss of the NO ligand. An over damped low
frequency bath mode is also included to broaden the spectral
function. In the panels directly below the lineshape, the time
dependent population transfer $P_{ef}(t)$ and the first moment
dynamics $\overline{Q}_{f}(t)$ in the product state are plotted.
$\overline{Q}_{f}(t)$ is obtained both by direct integration using
Eq. (\ref{b6}) and the initial first moments using Eqs.
(\ref{b9}-\ref{b11}). A reaction velocity of $\beta =2000 \
cm^{-1}/ps$ and $\beta=10000 \ cm^{-1}/ps$ are considered in
Figs.\ 8(b-e). The initial vertical energy, corresponding to the
separation of the quintet and singlet energy levels in MbNO, is
taken to be $\Omega _{00}^{ef}(0)=2000 \ cm^{-1}$. The connection
between the time dependent population evolution and the first
moment dynamics in the product state is clearly exposed in the
simulations. We see that for smaller $\beta $, the reaction
proceeds in a smooth fashion, with a complete transition to the
product occurring immediately after $t=1ps$. The slow passage
through the reaction vertex causes the product state oscillations
to be effectively diminished. For large $\beta $, the product
electronic state population changes more abruptly, and the initial
amplitude of the oscillatory motion is considerably larger,
approaching the equilibrium displacement of the reactant well with
respect to the product ($\left| \Delta _{ef}\right| =1$).

In Figs. 8(g-h), we simulate the amplitude and phase profiles of
the pump-probe signals associated with the reaction driven
oscillations in the electronic state $\left| f\right\rangle $,
using the inhomogeneous equilibrium lineshape of Mb shown in Fig.\
8(f). Note that only the product state electronic transition
$f\rightarrow f'$ is inhomogeneosly distributed. The probe pulse
detects the modulation of the $f \rightarrow f'$ electronic
transition by the reaction driven oscillations. Since the initial
conditions of the wave-packet on the state $\left| f\right\rangle
$ are independent of the pump wavelength, a sharp jump in the
phase is observed accompanied by a dip in the amplitude at the
absorption maximum $\Omega _{v}^{f}$. The direct correspondence
between the amplitude of the vibrational oscillations and the
amplitude of the signal is evident from the simulations. The
slower reaction exhibits a much weaker signal amplitude than the
faster reaction. The signal phase directly yields the initial
phase $\varphi _{f}$ of the wave-packet according to Eq.
(\ref{68cc}). Furthermore, if we let $\gamma =0$ and
$\stackrel{\cdot}{P}_{ef}(t)=\delta (t-\tau _{R})$ in Eqs.
(\ref{b10}) and (\ref{b11}), we have $\varphi _{f}\cong -\omega
_{0}\tau _{R}$, $\tau _{R}=\Omega_{00}^{ef}(0)/\beta $ being the
reaction time. Thus, when $\beta $ is increased, $\tau _{R}$\
decreases and the initial phase continually changes through cycles
of $2\pi $. The phase approaches zero for an infinitely fast
reaction. It is interesting to note from Fig.\ 8(c) and (e) that
the effective initial momentum of the wave-packet (due to a
non-zero $\varphi _{f}$) reverses its sign as $\beta $ is
increased from $2000 \ cm^{-1}/ps$ to $10000 \ cm^{-1}/ps$.
Compare Fig.\ 3(c), where the momentum impulse has the same
direction across resonance.

As an example, we consider the Fe-His mode in the reactive sample
MbNO in Fig.\ 7(c). In contrast to the non-reactive Mb, the phase
of the 220$ \ cm^{-1}$ oscillation in MbNO exhibits a sharp jump
at a wavelength that is red shifted with respect to the Soret
absorption maximum at $\sim 23000 \ cm^{-1}$\ $(435 \ nm)$ of the
Mb product. Also, the amplitude of the $220 \ cm^{-1}$
oscillations vanishes at this detection wavelength\cite{rosca00}.
These observations are consistent with the suggestion that the
Fe-His mode is driven by the chemical reaction associated with the
$e\rightarrow f$ electronic transition as described in Fig.\ 8.
The $4 \ nm$ red shift of the phase flip is probably associated
with the initial red shift of the photo-product absorption
band\cite{rosca00} centered roughly at $439 \ nm$ $(\sim 22780 \
cm^{-1}$). The fact that the phase flip occurs precisely at this
wavelength is consistent with the idea that the 220$ \ cm^{-1}$
oscillations in MbNO are associated with the initial Mb
photo-product that is created following photodissociation of the
NO ligand.

It is also seen from Fig.\ 7(c) that the phase of the open band
220$ \ cm^{-1}$ oscillations is near $0$ or $\pi $. This suggests
that the initial phase of the 220$ \ cm^{-1}$ mode coherence is
also close to $0$. A zero initial wavepacket phase is realized if
the reaction time $\tau _{R}$ is a small fraction of the $150\ fs
$ oscillatory period. The wave-packet is then placed almost
instantly ($\lesssim 20\ fs$) on the product state potential
surface. In the Landau-Zener theory, the reaction time depends
both on the reaction velocity $\beta$ and on the initial vertical
energy gap $\Omega_{00}^{ef}(R(0))$, suggesting that either the
gap is small or the velocity is large (or both) in the MbNO
reaction. As discussed earlier, this corresponds to a population
evolution that behaves like a step function.

On the other hand, a zero initial phase of the wavepacket could
also be an indication of direct pump pulse preparation of the
product state coherence. For example, when $\Omega
_{00}^{ef}(R(0))$ is very small, the pump pulse interaction may
directly excite an adiabatic mixture of the $\left| e\right\rangle
$ and the $\left| f\right\rangle $ electronic states. The
coherence induced along the $Q$ coordinate will then have a phase
of zero (no initial momentum), since it originated from the
excited state directly prepared by the pump pulse. This
possibility is beyond the scope of the present treatment, which
assumes that the pump pulse and chemical reaction can be
separated. We will address this interesting question in future
work.

Finally, we point out that the absolute phase of the 220$ \
cm^{-1}$ mode in MbNO indicates that the Fe-His bond in Mb
($\left| f\right\rangle$) contracts upon optical excitation to
Mb$^{*} \ (\left| f'\right\rangle $), as depicted in Fig.\ 1(b).
This is based on the knowledge that the Fe-His bond also contracts
upon loss of the NO ligand; i.e. upon the non-radiative transition
between $\left| e\right\rangle $ and $\left| f\right\rangle $. The
wavepacket induced in $\left| f\right\rangle $ is therefore placed
at positive displacements with respect to the equilibrium
position. It is clear from Fig.\ 1(b) that if the Fe-His bond
expands upon photo-excitation to $\left|f^{\prime}\right\rangle$,
exactly the opposite phase behavior from that observed in Fig.\
7(c) would be predicted. This confirms earlier
predictions\cite{stavrov93} and also demonstrates that the phase
of reaction driven pump-probe signals can provide information on
the sign of the optical electron nuclear coupling, provided the
initial ground state equilibrium changes are known.
\begin{figure}[htbp]
\begin{center}
\mbox{\epsfig{file=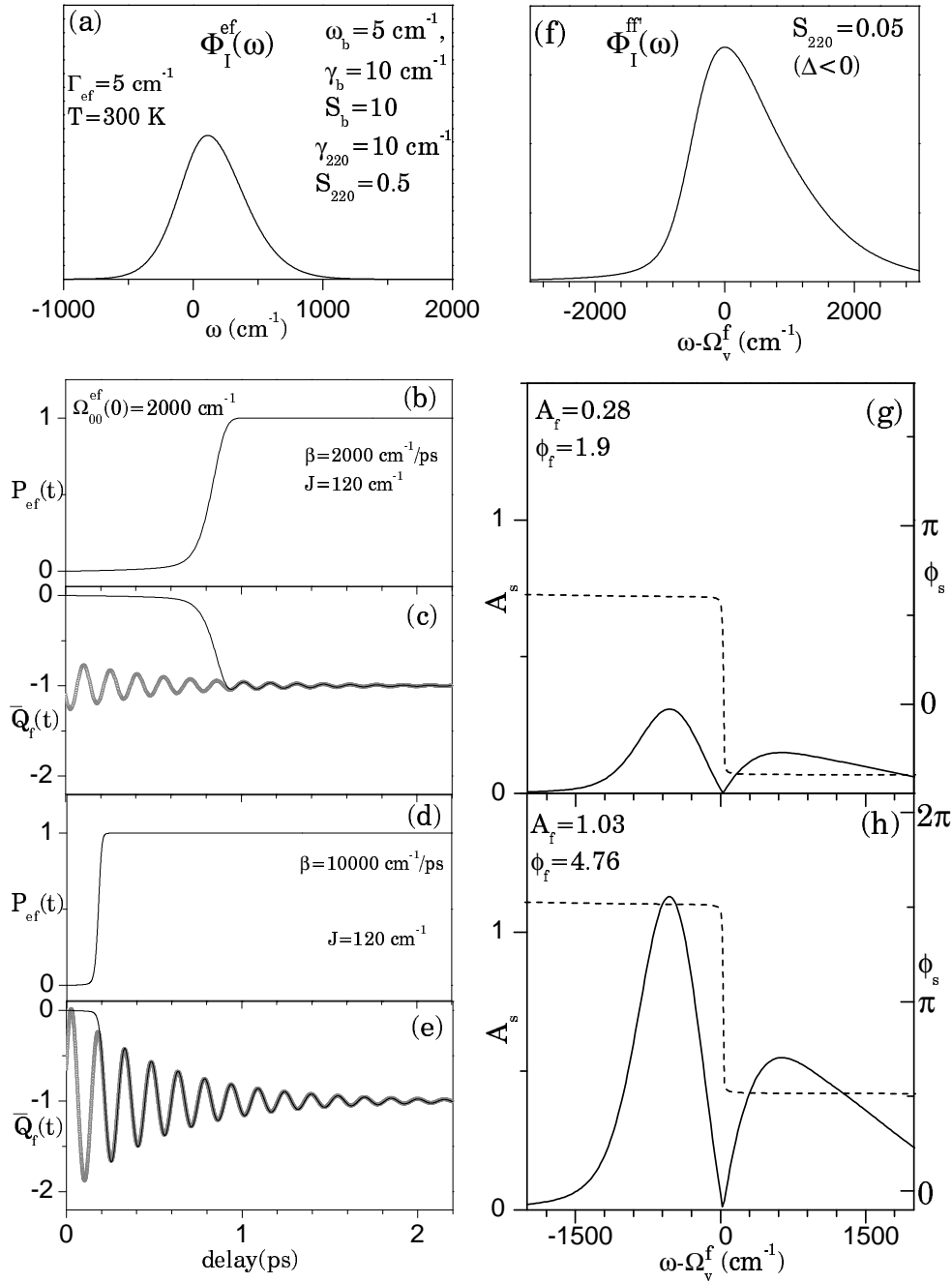,height=180mm}} \caption{Application
of the effective linear response approach to reaction driven
coherence. (a) Spectral lineshape function for the non-radiative
$e\rightarrow f$ electronic transition. Panels (b) and (d) show
the time dependent population evolution of the product state for
reaction velocities of $\beta =2000 \ cm^{-1}/ps$ and $\beta
=10000 \ cm^{-1}/ps$. Panels (c) and (e) show the corresponding
mean position of the $220 \ cm^{-1}$ mode coupled to the reaction.
The calculations using the effective initial conditions Eqs.
(\ref{b9}-\ref{b11}) are shown in circles. The results of a direct
integration as in Eq. (\ref{b6}) are shown as a solid line. The
non-radiative coupling strength was fixed at $J=120 \ cm^{-1}$ in
both the calculations. (f) Spectral lineshape function for the
product state (identical to the Mb lineshape in Fig.\ 7(a)). (g)
and (h) show the pump-probe open band amplitude $A_{s}$ (solid
line) and phase $\phi _{s}$ (dashed line) profiles arising from
the reaction driven $220 \ cm^{-1}$ coherence in the product
state, for $\beta =2000 \ cm^{-1}/ps$ and $\beta=10000 \
cm^{-1}/ps$.}
\end{center}
\end{figure}
\subsection{Overtone signals}

\subsubsection{First moment modulation (Coherent state)}

Up to this point, we have focused on the oscillatory pump-probe
signal at the fundamental frequency of the optically coupled mode.
The presence of higher overtones is expected even for harmonic
modes, since the observed signal is not directly the mean nuclear
position oscillating at the fundamental frequency. The pump-probe
signal arises from a dynamic modulation of the equilibrium
correlator as expressed in Eq. (\ref{50}) and (\ref{54}).
Intuitively, if we picture a two electronic level system, the
overtone signal can be attributed to the ``double pass'' of the
wave-packet across the probing region\cite{vos93,jonas95}. The
strength of the overtone signals due to the double pass is
determined by the higher powers of the product $A_{g}\Delta$. This
can been seen by expanding the exponential in the correlation
function in Eq. (\ref{49}). For weak field intensities, the pump
induced ground state amplitudes are much smaller than the optical
coupling strengths\cite{banin94,smith96,kumar00}, which are often
much less than unity in large multimode systems. Thus, the
overtones generated by the first moment modulation of the pump
induced ground state wave-packet are much weaker than the
fundamental.

On the other hand, when the coherence is induced by a
non-radiative surface crossing or by intense pump
fields\cite{banin94,smith96}, the amplitude of the coherent motion
can be comparable or larger than the optical coupling strengths,
so that the strength of the overtones is expected to be larger.
The effective linear response approach using displaced
wave-packets allows a calculation of pump-probe signals for
arbitrary values of the initial amplitude $A_{g}$. In order to
estimate the strength of overtone signals as a function of
$A_{g}$, we consider the model lineshape of Mb in Fig.\ 8(f). The
overtone signals of a low frequency mode at $40 \ cm^{-1}$ coupled
to the optical transition are studied in Fig.\ 9. The top panel of
Fig.\ 9 shows the profiles of the fundamental and the first two
overtones of the 40$ \ cm^{-1}$ oscillation, assuming $S_{\Delta
}=\Delta ^{2}/2=1$, and $S_{a}=A_{g}^{2}/2=10$. The contrasting
behavior of the overtone and fundamental amplitude profiles is
notable from the figure. The fundamental signal amplitude dips
near the resonant maximum $\Omega _{v}$ and peaks near the
shoulders of the absorption spectrum. The amplitude of the first
overtone peaks near $\Omega _{v}$ and vanishes close to the
frequency where the fundamental dips. Thus, we can say that the
amplitude profiles of the fundamental and the first overtone
behave roughly as the first and second derivatives of the
equilibrium absorption spectrum. Similarly, the second overtone at
120$ \ cm^{-1}$ behaves as the third derivative of the absorption
spectrum, vanishing at three different carrier frequencies across
the absorption band. Such characteristics are useful for the
assignment of overtone oscillations in pump-probe
spectroscopy\cite{rosca00,vos93}.

It is important to note the significant reduction in the amplitude
of the overtone oscillations compared to the fundamental. To study
the strength of overtone signals over a wider range of initial
displacements, we plot in Figs. 9(b-d), the amplitude of the first
three harmonics for three different carrier frequencies near
resonance. The initial wave-packet amplitudes are varied over
three orders of magnitude and $S_{\Delta }$ is fixed to be unity.
It is seen that for detection at the wings of the absorption
$\omega_{c}=\omega _{c1}$ and $\omega _{c3}$, the amplitude of the
overtones is more than an order of magnitude weaker than the
fundamental. This is true even for amplitudes as large as
$S_{a}=100$ (or $A_{g}\approx 14$). Near band center $\omega
_{c2}$, the fundamental signal itself is very small. The first
overtone is the dominating contribution even for very small
displacements. The almost linear behavior of the overtone
amplitudes on the log-log scale suggests a power law dependence of
the overtone amplitudes on the wavepacket displacement.

\subsubsection{Second moment modulation (Squeezed states)}

While the contribution of the double pass of the wave-packet to
the intensity of overtone signals is negligible for field driven
coherence (see previous section), higher moment modulations of the
pump induced wave-packet\cite{kumar00} (squeezing) can contribute
more significantly to overtone intensities. Overtones signals
calculated using the full third order response approach are found
to be much stronger than those calculated with the effective
linear response approach using the first moment dynamics. This
discrepancy is indicative of the role played by field induced
squeezing in the ground state signals. Apart from the laser pulse
interaction, squeezing of the nuclear wave-packet can also be
induced by purely geometrical effects, such as a change in the
curvature of the nuclear potentials that are involved in the
interaction\cite{janszky94} i.e. quadratic coupling.

We discuss squeezing in the context of chemical reactions.
Introducing quadratic coupling between the potentials of the
reactant and product surfaces $\left|e\right\rangle $ and $\left|
f\right\rangle $ shown in Fig. 1, we calculate the probe
response to displaced and squeezed dynamics on the product state
potential well. For the initial displacement, we continue to use
the first moments of the Landau-Zener coherence presented in
Appendix B. These were derived assuming linear coupling between
the reactant and product surfaces. To obtain the initial
conditions of the squeezed state, we adopt a phenomenological
view. We assume that the effect of the reaction is to instantly
place the thermal equilibrium distribution of the reactant on the
product state well. The resulting non-stationary nuclear state is
a displaced-thermal density matrix that is also squeezed due to
the incommensurate thermal widths on the two surfaces. In
Appendix-D, we have derived an analytic expression for the
effective linear response correlation function that describes both
the coherent and squeezed dynamics in the ground state.

The central quantity governing the squeezed state dynamics is the
ratio of the reactant and product state frequencies
$r=(\omega_{e}/\omega _{f})$. It is evident from the effective
linear response correlation function Eq. (\ref{y11}) that squeezed
states modulate the equilibrium absorption correlator at even
harmonics of the fundamental frequency. We therefore do not expect
the fundamental pump-probe signal to be affected significantly by
squeezed wave-packet dynamics. In Figs. 9(e-g), we simulate the
first three harmonics of the $40\ cm^{-1}$ oscillations, when the
wave-packet is both displaced and squeezed (The amplitude and
phase profiles in the presence of squeezing behave similar to
those in Fig.\ 9(a) and are not plotted). In the simulations, the
initial displacement of the wave-packet on the $\left|
f\right\rangle $ state and the optical coupling are fixed at
$S_{a}=S_{\Delta }=1$. The squeezing ratio $r$ is varied over a
range of $1$ to $8$. It can be seen that fundamental oscillations
are practically unaffected by variations in $r$. The dramatic
effect of squeezing on the overtone amplitude is evident,
particularly so for the detection frequency $\omega_{c2}$ near the
absorption maximum. Here, for squeezing ratios larger than 6, the
overtone signal is comparable to the corresponding fundamental
signal at the wings of the absorption band ($\omega _{c1},\omega
_{c3}$).
\begin{figure}[htbp]
\begin{center}
\mbox{\epsfig{file=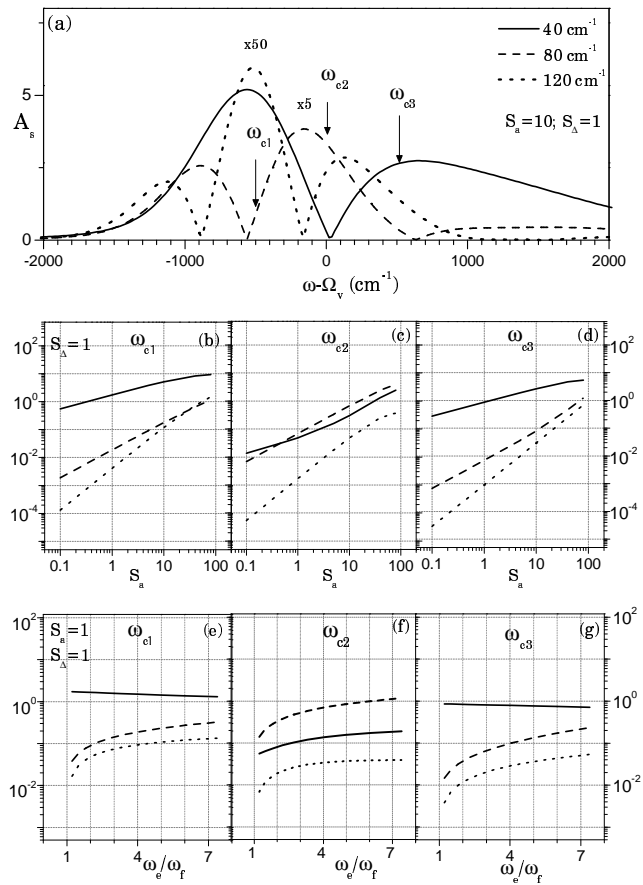,height=120mm}} \caption{Simulations
of overtone signals from the ground state of Mb using displaced
and squeezed states. (a) Amplitude profiles of the fundamental at
$40 \ cm^{-1}$ (solid line) and the first two overtones at $80 \
cm^{-1}$ (dashed line) and $120 \ cm^{-1}\,$(dotted line),
assuming an initial displacement of $S_{a}=A_g^2/2=10$, and
optical coupling strength of $S_{\Delta }=\Delta^2/2=1$, for the
$40 \ cm^{-1}$ mode. Panels (b-d) show the amplitudes of the first
three harmonics over a wider range of initial displacements, for
three carrier frequencies ($\omega _{c1},\omega _{c2},\omega
_{c3}$) across the absorption maximum. The corresponding
amplitudes of the three harmonics for displaced and squeezed
states is plotted in panels (e-g), for a wide range of frequency
ratios $r=\omega _{e}/\omega _{f}$ between the reactant
($\left|e\right>$) and product ($\left|f\right>$) electronic
states.}
\end{center}
\end{figure}
The simulations presented in Fig.\ 9 provide an overall picture of
overtone intensities in pump-probe spectroscopy, within a harmonic
model with coherent and squeezed initial states. It can be
concluded that the fundamental signal is almost always the
dominant contribution for reasonable values of wave-packet
displacement and squeezing. However, when quadratic coupling is
also introduced between the electronic levels involved in the
probe interaction, i.e. when $\omega_{f}\neq \omega _{f^{\prime
}}$, the overtone signals are expected to be significantly
enhanced. In this case, the strength of the overtone oscillations
induced by the curvature change (between the ground and excited
state potentials) is coupled to the amplitude of the wave-packet
motion; larger amplitudes imply that steeper differences in
curvature are probed. In contrast, the overtone intensities
generated by squeezing the initial wave-packet are dependent only
on the frequency ratio. A rigorous treatment of the effective
linear response functions for a quadratically coupled two level
system will be presented elsewhere.

\section{Summary and conclusions}

In this work, we have presented a study of the amplitude and phase
excitation profiles in femtosecond coherence spectroscopy using an
effective linear response approach. In combination with a rigorous
moment analysis of the doorway state\cite{kumar00}, the linear
response formalism is found to be in excellent agreement with the
more complete third order response formalism\cite{muk95}.
Furthermore, the representation of non-stationary states as
displaced thermal states provides a general framework to analyze
vibrational coherence induced by non-radiative transitions. We
have demonstrated this approach by using the first moments of a
quantum coordinate coupled to a Landau-Zener surface crossing.

The effective linear response functions are analytical, and yet
retain the correlation function based description (in contrast to
the vibronic eigenstates approach\cite{fain92,fain93b}) implicit
in the third order response formalism. The analytic expressions
offer a significant reduction in computation times. Calculations
of multiple carrier frequency excitation profiles for arbitrary
pulse widths and temperatures are made possible, without the need
for semi-classical approximations. A separate moment analysis is
also useful in providing clear physical insight into pump-probe
signals. Ground state signals generated by impulsive Raman
processes exhibit contrasting phase and amplitude behaviour from
excited state and reaction driven signals.

The ability to incorporate experimentally measured absorption
lineshapes (and the Kramers-Kronig determined dispersion
lineshapes) into the pump-probe calculations is another important
aspect of the present work. A knowledge of pump-probe excitation
profiles and the measured absorption cross-sections can be used to
extract parameters relevant to a specific mode that is active in
pump-probe signals. The use of experimental lineshapes eliminates
the need for modeling the multimode vibronic mechanisms that cause
line broadening. This is analogous to ``transform'' techniques in
resonance Raman
spectroscopy\cite{page81,Stallard83,Schomacker89,page91}.
Transform techniques are a powerful means for extracting mode
specific information using the measured Raman excitation profiles
and the optical absorption profiles on an absolute (cross-section)
scale.

A study of inhomogeneous broadening in pump-probe signals shows
that the ground state amplitude and phase profiles are smeared out
by the inhomogeneous broadening. The excited state amplitude dip
and phase flip are red shifted with respect to the peak of the
emission lineshape. A pump-probe analysis of the phase profiles of
the 220$ \ cm^{-1}$ Fe-His mode in Mb across the Soret band was
presented, and was based on a previous model for the strongly
asymmetric Soret lineshape\cite{srajer86}. The analysis indicates
both the ground state origin of the oscillations and the strong
inhomogeneous broadening present in Mb. In contrast to the nearly
constant phase profile in Mb, the 220$ \ cm^{-1}$ mode phase in
the reactive sample MbNO shows a clear $\pi$ phase change near the
peak of the transient photo-product. These results suggest that
the 220$ \ cm^{-1}$ mode coherence in MbNO arises from the
photo-product Mb. The mode is presumably triggered by the rapid
curve crossing occurring during the MbNO photolysis
reaction\cite{ZHU94}. The absolute phase of the oscillations
suggest that the chemical reaction occurs on a very fast time
scale ($\lesssim 20\ fs$), with the possibility of direct pump
pulse involvement at the crossing region in the preparation of the
product state coherence.

Although we have not discussed vibrational relaxation in detail,
the non-stationary linear response correlation functions in Eqs.
(\ref{50}) and (\ref{54}) can readily incorporate damped
wave-packet motion. The initial first moments for the damped
oscillator case must however be used instead of the undamped
expressions in Appendix-A. In the analysis of the oscillatory
amplitude and phase profiles, which has been the main subject of
the present work, the neglect of damping is justified because the
observed modes are weakly damped. However, both under damped and
over damped modes contribute to the non-oscillatory background
(offset) in the pump-probe signals\cite{smith94,jonas95,yang99}.
These offsets can potentially carry important information
concerning the equilibrium potential shifts\cite{yang99}. It is
immediately clear from Eqs. (\ref{50}) and (\ref{54}) that the
modulation of the equilibrium correlator by strongly over damped
or diffusive motion manifests itself as a dynamic i.e. shifting
absorption lineshape in the ground state response\cite{demidov00}.
Overdamped motion is manifested in the excited state response
through phenomena such as fluorescence Stokes-shift
dynamics\cite{muk95,ungar97}. A rigorous analysis of these
processes can be readily carried out using the effective linear
response approach formalism presented here.

We have also discussed overtone signals in pump-probe spectroscopy
using both displaced and squeezed initial states. The contrasting
amplitude and phase behavior of the various harmonics are
potentially useful for overtone assignments. It is found that for
reasonable values of squeezing ratios and wave-packet
displacements, the fundamental signal is the dominant
contribution.

Finally, we mention that the present development can be readily
extended beyond the displaced oscillator model. Effective linear
response functions can be readily calculated for the case of
quadratic electron-nuclear coupling (between the radiatively
coupled levels) using standard techniques of quantum field theory.
Non-Condon effects can also be
incorporated\cite{tanimura93,khidekel96} into both the pump
(moment analysis) and the probe steps. These calculations will be
detailed elsewhere.

\begin{center}
{\bf Acknowledgement}
\end{center}
We thank Dr. T. S. Yang for helpful comments regarding the
numerical implementation of the third order response approach.
This work was supported by the National Science Foundation (MCB
9904516) and by the National Institutes of Health (AM 35090).

\newpage
\appendix

\section{First moments of pump induced vibrational coherence}

Consider the ground state doorway $\delta \widehat{\rho }_{g}$ in
Eq. (\ref{26}). Let the pump field be of the form
$E_{a}(t)=E_{0}G(t)\cos (\omega _{c}t)$ where $G(t)\,$\ is a
dimensionless envelope function, and $E_{0}\,$is the electric
field strength of the pump pulse. General expressions have been
derived for all the moments of $\widehat{Q}$ and $\widehat{P}$\
for the density matrix $\delta \widehat{\rho }_{g}$ using moment
generating functions, and related to equilibrium lineshape
functions\cite{kumar00}. The first moments of $\widehat{Q}$ and
$\widehat{P}$ are found to be (in dimensionless units):
\begin{eqnarray}
Q_{g0} &=&-
{\displaystyle
{\left| \mu _{ge}\right| ^{2}E_{0}^{2}(2\overline{n}+1)\Delta
\over 8\pi \hbar ^{2}N_{g}^{\prime }}}
\int_{0}^{\infty }
d\omega \widetilde{G}_{p}(\omega -\omega _{c},-\omega _{0})
\widehat{\Delta }\Phi _{I}(\omega ),  \label{41a} \\
P_{g0} &=&
{\displaystyle {\left| \mu _{ge}\right| ^{2}E_{0}^{2}\Delta
\over 8\pi \hbar ^{2}N_{g}^{\prime }}}
\int_{0}^{\infty }d\omega
\widetilde{G}_{p}(\omega -\omega _{c},-\omega _{0})
\widehat{\Delta }\Phi _{R}(\omega ).  \label{41b}
\end{eqnarray}
Here, $N_{g}^{\prime }$ is the net ground state population after
the pump interaction and $\Phi _{I}(\omega )$ and $\Phi
_{R}(\omega )$ are the imaginary and the real parts of the complex
equilibrium lineshape function defined in Eq.(\ref{41d}).
$\widetilde{G}_{p}(\omega -\omega _{c},-\omega _{0})$ is the
product spectral function defined in Eq. (\ref{65cc}). The action
of the operator $\widehat{\Delta }\,\ $(not to be confused with
the dimensionless excited state potential shift $\Delta $) is to
generate differences: $\widehat{\Delta }\,\Phi _{I}(\omega )=\Phi
_{I}(\omega )-\Phi _{I} (\omega -\omega_{0}).$ With the initial
position and momentum increments given in Eqs. (\ref{41a}) and
(\ref{41b}), the time dependent first moment dynamics in the
ground state is given rigorously as the solution to the damped
oscillator equation:
\begin{equation}
\overline{Q}_{g}(s)=A_{g}e^{-\gamma \left| s\right| }\cos (\omega
_{v}s+\varphi _{g}),  \label{47}
\end{equation}
where $\gamma $ and $\omega _{v}=\sqrt{\omega _{0}^{2}-\gamma
^{2}}$ are the damping constant and effective frequency of the
mode. The amplitude and phase are given by
$A_{g}=\sqrt{Q_{g0}^{2}+P_{g0}^{2}}$ and $\varphi_{g}=-\tan
^{-1}\left[ P_{g0}/Q_{g0}\right] $.

The various moments of $\widehat{Q}\,$and $\widehat{P}$ for the
excited state doorway $\delta \widehat{\rho }_{e}$ in Eq.
(\ref{33}) can also be obtained. It is found\cite{kumar00} that
the excited nuclei receive no initial momentum i.e., $P_{e0}=0$.
The mean position (with respect to the excited state potential
equilibrium $\Delta$) is given by:
\begin{eqnarray}
Q_{e0} &=&-{\displaystyle {\left| \mu _{ge}\right| ^{2}E_{0}^{2}
\Delta \over 4\pi \hbar ^{2}N_{e}^{\prime }}}
\int_{0}^{\infty }d\omega \widetilde{G}_{p}(\omega -\omega _{c},
-\omega _{0})\nonumber \\
&&\times \left[ \Phi _{I}(\omega -\omega _{0})-\overline{n}
\widehat{\Delta }\Phi _{I}(\omega )\right] ,  \label{47a}
\end{eqnarray}
where $N_{e}^{\prime }$ is the electronic population in the
excited state ($N_{g}^{\prime }+N_{e}^{\prime }=1$). Thus, the
amplitude for the excited state coherent motion is given by
$A_{e}=\left| Q_{e0}\right| $, and the time dependent first moment
is
\begin{equation}
\overline{Q}_{e}(s)=A_{e}e^{-\gamma \left| s\right| }\cos (\omega _{v}s).
\label{47b}
\end{equation}

\section{First moments of coherence driven by Landau-Zener surface crossing}

We consider a product electronic state $\left| f\right\rangle $
that is coupled non-radiatively to the reactant excited state
$\left|e\right\rangle $ as in Fig.\ 1(b). The Hamiltonian of the
problem consists of the $2\times2$ lower block diagonal matrix of
$\widehat{H}_{NR}$ in Eq. (\ref{68a}). The system is assumed to be
initially ($t=0$) in thermal equilibrium along the quantum
mechanical degrees of freedom of the state $\left| e\right\rangle
$. The non-radiative coupling $J$ will induce a transition to the
state $\left| f\right\rangle $. The time-dependent surface
crossing along the classical reaction coordinate $R$ will be
accompanied by quantum mechanical tunneling between the
vibrational levels of $\widehat{H}_{e}$ and $\widehat{H}_{f}$. The
time dependent quantum yield for the transition $e\rightarrow f$
can be shown to be:
\begin{equation}
P_{ef}(t)=1-\exp \left\{ -\int_{0}^{t}ds{\cal
G}(R(s))\right\}.\label{b2}
\end{equation}
Here, ${\cal G}(R(s))$ is the Fermi Golden rule transition rate
for making an electronic transition from the state $\left|
e\right\rangle $ to the electronic state $\left| f\right\rangle $
and is expressed as\cite{zhu97}:
\begin{equation}
{\cal G}(R(s))={\displaystyle {2\pi J^{2} \over \hbar ^{2}}} \Phi
_{I}^{ef}(\omega =0,\Omega _{00}^{ef}(R(s))).  \label{b3}
\end{equation}
$\Omega _{00}^{ef}(R(s))=U_{f}(R(s))-U_{e}(R(s))$ is the
time-dependent energy gap between the $e$ and $f$ states.
$\Phi_{I}^{ef}(\omega )$ is the spectral absorption lineshape
function for the $e\rightarrow f$ electronic transition, obtained
using Eqs. (\ref{41}-\ref{41d}) with $\Omega _{00}$ replaced by
$\Omega _{00}^{ef}(R(s))$. In the single mode limit, i.e. no $Q$
coupling, Eq. (\ref{b3}) yields a delta function. The time
dependent quantum yield $P_{ef}(t)$ will then be a simple step
function in time. The presence of additional degrees of freedom
smears out the curve crossing process and can be completely
described by the zero frequency component of the spectral
lineshape function $\Phi _{I}^{ef}(\omega )$.

The time dependent surface crossing along the $R$ coordinate
induces oscillations along $Q$, which is initially in thermal
equilibrium. It can be shown generally that the mean position of
the oscillator in the product state obeys\cite{zhu97}:
\begin{equation}
\overline{Q}_{f}(t)= {\displaystyle {\omega _{0}^{2}\Delta _{ef}
\over \omega _{v}}} \int_{0}^{\infty }dse^{-\gamma s}\sin (\omega
_{v}s)P_{ef}(t-s). \label{b6}
\end{equation}
Here, $\gamma $ is the vibrational damping, $\Delta _{ef}$ is the
dimensionless shift between the $e$ and $f$ oscillators, and
$\omega _{v}=\sqrt{\omega _{0}^{2}-\gamma ^{2}}$. The above
equation rigorously describes the first moment of the nuclear
dynamics on the product state well, with the one qualification
that it is valid only at times $t$\ when the system has made its
transition into the product state, i.e. $P_{ef}(t)=1$. When
$P_{ef}(t)<1$, $\overline{Q}_{f}(t)$ consists of oscillations in
both the reactant and product surfaces and is more difficult to
interpret. We can recast Eq. (\ref{b6}) to a form that readily
yields the initial conditions for the probe i.e. detection step.
Integrating Eq. (\ref{b6}) by parts, we get in the long time limit
(assuming $P_{ef}(\infty )=1$)
\begin{equation}
\overline{Q}_{f}(t)=\Delta _{ef}+A_{f}e^{-\gamma t}\cos (\omega
_{v}t+\varphi _{f}),  \label{b9}
\end{equation}
where
\begin{eqnarray}
A_{f}\cos (\varphi _{f}) &=&-\Delta _{ef}\int_{-\infty }^{\infty }dt^{\prime
}\stackrel{.}{P}_{ef}(t^{\prime })e^{\gamma t^{\prime }}\left[ \cos (\omega
_{v}t^{\prime })-(\gamma /\omega _{v})\sin (\omega _{v}t^{\prime })\right] ,
\label{b10} \\
A_{f}\sin (\varphi _{f}) &=&\Delta _{ef}\int_{-\infty }^{\infty }dt^{\prime }
\stackrel{.}{P}_{ef}(t^{\prime })e^{\gamma t^{\prime }}\left[ \sin (\omega
_{v}t^{\prime })+(\gamma /\omega _{v})\cos (\omega _{v}t^{\prime })\right].
\label{b11}
\end{eqnarray}
Eqs. (\ref{b10}) and (\ref{b11}) determine the amplitude and phase
of the oscillatory motion induced by the surface crossing. They
can be readily incorporated into the probe step using a displaced
thermal representation. The above expressions depend on the
reaction parameters ($J,\beta ,\Omega _{00}^{ef}(R)$, etc.,)
through $P_{ef}(t)$ and Eq. (\ref{b2}). Extension of the
multi-dimensional Landau-Zener theory to incorporate
non-stationary initial conditions along the $Q$ degree of freedom
can be carried out using displaced thermal density matrices, with
 the displacements obtained from the excited state first moment
presented in Appendix A.

\section{Non-linear response functions}

When the ground state doorway $\delta \widehat{\rho} _{g}$ in
 Eq. (\ref{26}) is substituted in the expression for
$C_{g}(t,t_{3})$ in Eq. (\ref{25}), we are left with the thermal
average of a four-time correlation function. The average can be
evaluated exactly using a second order cumulant expansion and the
result is
\begin{equation}
C_{g}(t,t_{3})=-\frac{\mu ^{2}}{\hbar ^{2}}\int_{-\infty }^{\infty
}dt_{2}\int_{-\infty }^{t_{2}}dt_{1}E_{a}(t_{1})E_{a}(t_{2})\left[
R_{4}(t,t_{3},t_{2},t_{1})+R_{3}(t,t_{3},t_{2},t_{1})\right].
\label{27}
\end{equation}
Here, $R_{3}$ and $R_{4}$ are the non-linear response functions
given by\cite{foot4,muk82,muk95},
\begin{mathletters}
\begin{eqnarray}
R_{3}(t,t_{1},t_{2},t_{3}) &=&{\cal H}(t-t_{3}){\cal H}(t_{2}-t_{1})
e^{-i\Omega_{00}(t-t_{3}-t_{2}+t_{1})}e^{-g(t-t_{3})-
g^{*}(t_{2}-t_{1})+f_{2}^{*}(t,t_{1},t_{2},t_{3})},\label{28} \\
R_{4}(t,t_{1},t_{2},t_{3}) &=&{\cal H}(t-t_{3}){\cal H}(t_{2}-t_{1})
e^{-i\Omega_{00}(t-t_{3}+t_{2}-t_{1})}e^{-g(t-t_{3})-
g(t_{2}-t_{1})-f_{2}(t,t_{1},t_{2},t_{3})},
\label{29}
\end{eqnarray}
\end{mathletters}
where we have defined
\begin{equation}
f_{2}(t,t_{1},t_{2},t_{3})=g(t-t_{1})-g(t-t_{2})+g(t_{3}-t_{2})-g(t_{3}-t_{1}).
\label{30}
\end{equation}
The function ${\cal H}(t)$ accounts for homogeneous broadening and
is typically of the form $e^{-\Gamma _{e}\left| t\right|}$,
$\Gamma _{e}$ being the homogeneous damping constant. In these
expressions, $t_{2}-t_{1}$ reflects the time interval when the
system is in an electronic coherence during the pump field
interaction, whereas $t-t_{3}$ is the time interval when the
system is in an electronic coherence during the probe
interaction\cite{muk95}. Thus, ${\cal H}(t)$ appears separately
for the pump and the probe interactions.

Turning to the excited state response, substitution of $\delta
\widehat{\rho} _{e}$ in Eq. (\ref{33}) into the expression for
$C_{e}$ in Eq. (\ref{32a}) and using a second order cumulant
expansion, we find
\begin{equation}
C_{e}(t,t_{3})=\frac{\mu ^{2}}{\hbar ^{2}}\int_{-\infty }^{\infty
}dt_{2}\int_{-\infty }^{t_{2}}dt_{1}E_{a}(t_{1})E_{a}(t_{2})\left[
R_{1}^{*}(t,t_{1},t_{2},t_{3})+R_{2}^{*}(t,t_{1},t_{2},t_{3})\right].
\label{34}
\end{equation}
Here, $R_{1}$ and $R_{2}$ are the non-linear response functions given
by\cite{muk82,muk95}
\begin{mathletters}
\begin{eqnarray}
R_{1}(t,t_{1},t_{2},t_{3}) &=&{\cal H}(t-t_{3}){\cal H}(t_{2}-t_{1})
e^{-i\Omega_{00}(t-t_{3}+t_{2}-t_{1})}e^{-g^{*}(t-t_{3})-
g(t_{2}-t_{1})-f_{1}(t,t_{1},t_{2},t_{3})},\label{35} \\
R_{2}(t,t_{1},t_{2},t_{3}) &=&{\cal H}(t-t_{3}){\cal H}(t_{2}-t_{1})
e^{-i\Omega_{00}(t-t_{3}-t_{2}+t_{1})}e^{-g^{*}(t-t_{3})-
g^{*}(t_{2}-t_{1})+f_{1}^{*}(t,t_{1},t_{2},t_{3})}, \label{36}
\end{eqnarray}
\end{mathletters}
and we have defined
\begin{equation}
f_{1}(t,t_{1},t_{2},t_{3})=g(t-t_{1})-g^{*}(t-t_{2})-g(t_{3}-t_{1})
+g^{*}(t_{3}-t_{2}).
\label{37}
\end{equation}
In the above expressions $g(s)$ is the harmonic oscillator
correlation function [Eq. (\ref{38})].

Equations (\ref{27}) and (\ref{34}) express the two time effective
linear response correlation functions $C_{g,e}(t,t_{3})$ as as
convolution of the well known non-linear response
functions\cite{pollard92,muk95} $R_{j}$ $(j=1..4)$ with the two
pump field interactions. The convolution with the pump laser
fields implicitly takes into account the complete information
about the non-stationary state of the system expressed by the
density matrices $\delta \widehat{\rho }_{g}$ and $\delta
\widehat{\rho }_{e}$. It is worth noting here that both
$C_{g}(t,t_{3})$ and $C_{e}(t,t_{3})$ can be factored into
equilibrium and non-equilibrium parts as follows:
\begin{mathletters}
\begin{equation}
C_{g}(t,t_{3})=K_{g}(t-t_{3})C_{g}^{\prime }(t,t_{3});\text{ \ \ }%
C_{e}(t,t_{3})=K_{e}(t-t_{3})C_{e}^{\prime }(t,t_{3}),  \label{40}
\end{equation}
\end{mathletters}
where $K_{g}(t-t_{3})$ and $K_{e}(t-t_{3})$ are given by Eq. (\ref{41}).

Inhomogeneous broadening due to a static distribution ${\cal
F}_{I}(\Omega_{00})$ can be incorporated into the above
expressions. The correlation functions $C_{g}$ and $C_{e}$ must be
averaged with respect to the inhomogeneous distribution. The
electronic energy gap appears in the response functions $R_{j}$
through the oscillatory factors $e^{-i\Omega _{00}\left[
(t-t_{3})\pm(t_{2}-t_{1})\right] }$. The averaging over ${\cal
F}_{I}(\Omega _{00})$ thus involves a simple Fourier transform and
can be carried out independently of the time integrations. The
resulting inhomogeneous response functions $R_{j}^{(I)}$ can then
be expressed in terms of the homogeneous ones $R_{j}$ in the
following factorized form\cite{muk95}:
\begin{mathletters}
\begin{eqnarray}
R_{1,4}^{(I)}(t,t_{1},t_{2},t_{3}) &=&R_{1,4}(t,t_{1},t_{2},t_{3})
\widetilde{{\cal F}}_{I}(t-t_{3}+t_{2}-t_{1}),  \label{x3} \\
R_{2,3}^{(I)}(t,t_{1},t_{2},t_{3}) &=&R_{2,3}(t,t_{1},t_{2},t_{3})
\widetilde{{\cal F}}_{I}(t-t_{3}-t_{2}+t_{1}),  \label{x4}
\end{eqnarray}
\end{mathletters}
\smallskip where $\widetilde{{\cal F}}_{I}(t)$ is the Fourier transform of
the inhomogeneous distribution ${\cal F}_{I}(\Omega _{00})$.

\section{Effective linear response function for a two level system with
squeezed vibrational states}

Consider two non-radiatively coupled reactant
$\left|e\right\rangle $ and product $\left| f\right\rangle $
electronic states. Let the vibrational potentials be harmonic with
linear and quadratic coupling, with frequencies $\omega _{e}$ and
$\omega _{f}$. Since two different vibrational frequencies are
involved, we will first use dimensional units for clarity and will
revert to dimensionless units later. Just before the reaction, the
system is in thermal equilibrium on the electronic state
$\left|e\right\rangle$ with the coordinate and momentum
uncertainties ($\sigma_{A}^{2}=\left\langle
\widehat{A}^{2}\right\rangle - \left\langle
\widehat{A}\right\rangle ^{2}$) given by
\begin{equation}
\sigma _{Qe}^{2}=\left( \hbar /2m\omega _{e}\right) (2\overline{n}%
_{T}+1);\sigma _{Pe}^{2}=\left( m\omega _{e}\hbar /2\right) (2\overline{n}%
_{T}+1),  \label{y2}
\end{equation}
where $\overline{n}_{T}$ is the mean occupation number at
temperature $T$. The position and momentum uncertainties in the
product state are obtained from the above expression by simply
replacing $\omega _{e}$ by $\omega _{f}$. If we picture the
chemical reaction as instantly creating distributions of the above
widths in the product state well, the initial non-equilibrium
state in the product well will appear to be squeezed in the
position and momentum coordinates. The non-stationary state can be
generated from the thermal equilibrium density matrix of the
product well by the squeezing operator defined as\cite{mandel95}
\begin{equation}
\widehat{S}(\beta )=\exp \left[ \frac{\beta }{2}(\widehat{a}^{\dagger 2}-
\widehat{a}^{2})\right] ;\beta =-\frac{1}{2}\ln \left( \frac{\omega _{f}}
{\omega _{e}}\right).  \label{y3}
\end{equation}
The present choice of the squeezing operator maintains the minimum
uncertainty product. Arbitrary changes in the uncertainty product
can be incorporated\cite{mandel95} by a complex value for the
squeezing parameter $\beta$, but is not considered here for
simplicity. It is easy to verify that the unitary operator
$\widehat{S}(\beta )$ converts the thermal density matrix for the
state $\left| f\right\rangle $, i.e. $\widehat{\rho }_{T}^{(f)}$
to the thermal density matrix for the state $\left|e\right\rangle
$, i.e. $\widehat{\rho }_{T}^{(e)}$. The position and momentum
uncertainties are also changed to the thermal values in the
reactant state given in Eq. (\ref{y2}).

Since the wave packet induced in the state $\left| f\right\rangle
$ is in general both displaced and squeezed, we may represent the
corresponding density matrix as
\begin{equation}
\widehat{\rho }_{f}^{\prime }=\widehat{D}(\lambda _{f})\widehat{S}
(\beta )\widehat{\rho }_{T}^{(f)}\widehat{S}^{\dagger }(\beta )
\widehat{D}^{\dagger}(\lambda _{f}).  \label{y7}
\end{equation}
With the above representation, we evaluate the correlation function similar
to Eq. (\ref{25}). We find (reverting to dimensionless units where the
scaling factor is $\left( \hbar /m\omega _{f}\right) ^{1/2}$)
\begin{equation}
C_{f}{\large (}t{\large ,}t_{3})=e^{-i\Omega _{v}^{f}(t-t_{3})}Tr\left[
\widehat{\rho }_{T}\exp \left( i\omega _{f}\Delta \int_{t_{3}}^{t}ds
\widehat{S}^{\dagger }(\beta )\widehat{D}^{\dagger }(\lambda _{g})
\widehat{Q}(s)\widehat{D}(\lambda _{g})
\widehat{S}(\beta )\right) _{+}\right] ,  \label{y8}
\end{equation}
where we have used the invariance of trace under cyclic
permutation to bring the unitary operators inside the time ordered
exponential. Using the properties of coherent and squeezed
states\cite{mandel95}, we have
\begin{eqnarray}
&&\widehat{S}^{\dagger }(\beta )\widehat{D}^{\dagger }(\lambda
_{g}) \widehat{Q}(s)\widehat{D}^{\dagger }(\lambda
_{g})\widehat{S}(\beta ) \nonumber \\ &=&A_{f}\cos (\omega
_{f}s+\varphi _{f})+\left( \widehat{a} \Lambda
(s)+\widehat{a}^{\dagger }\Lambda ^{*}(s)\right),  \label{y9}
\end{eqnarray}
where $\Lambda (s)=\left( \cosh (\beta )e^{-i\omega _{f}s}+\sinh
(\beta)e^{i\omega _{f}s}\right)$. Substituting this in Eq.
(\ref{y8}), we can remove the c-number displacement outside the
time ordering. The resulting thermal average can be evaluated
exactly using Wick's theorem. We finally obtain
\begin{eqnarray}
C_{f}(t,t_{3}) &=&K_{f}(t-t_{3})\exp \left[ (iA_{f}\Delta \left( \sin
(\omega _{f}t+\varphi _{f})-\sin (\omega _{f}t_{3}+\varphi _{f})\right)
\right]  \nonumber \\
&&\times \exp \left[ -R\left( 2\cos (\omega _{f}(t+t_{3}))-\cos (2\omega
_{f}t)-\cos (2\omega _{f}t_{3})\right) \right] ,  \label{y11}
\end{eqnarray}
where we have defined
\begin{equation}
R=\frac{\Delta ^{2}}{2}(2\overline{n}_{T}+1)\cosh (\beta )\sinh
(\beta )={\displaystyle {(2\overline{n}_{T}+1)\Delta ^{2} \left(
r^{2}-1\right)  \over 8r}}.  \label{y12}
\end{equation}
Here, $r=\omega _{e}/\omega _{f}\,$ and $\Delta $ is the
dimensionless shift between the product ground and excited
potentials. The equilibrium correlator is given by
$K_{f}(t-t_{3})=e^{-i\Omega_{00}^{f}}e^{-g(t-t_{3})}$, where
$g(s)$ now takes a slightly different form as compared to Eq.
(\ref{38}):
\begin{equation}
g(t-t_{3})=\frac{\Delta _{f}^{2}}{2}\left[ (2\overline{n}_{T}^{\prime
}+1)\left( 1-\cos (\omega _{f}(t-t_{3}))\right) +i\sin (\omega
_{f}(t-t_{3}))\right] ,  \label{y13}
\end{equation}
where
\begin{equation}
(2\overline{n}_{T}^{\prime }+1)= \coth\left( {\hbar \omega
_{f}/}2k_{B}T\right) \left( \cosh ^{2}(\beta )+\sinh ^{2}\left(
\beta \right)\right). \label{y14}
\end{equation}
Eq. (\ref{y11}) expresses the non-equilibrium response as the
modulation of the equilibrium correlator $K_{f}(t-t_{3})$ by the
coherent and squeezed dynamics of the nuclear motion on the ground
electronic state $\left| f\right\rangle $ of the product. In
addition to the exponential factor that describes coherent
dynamics, squeezing introduces an additional factor that modulates
at the even harmonics of the fundamental oscillator frequency
$\omega _{f}$. It is seen that the strength of the overtone
modulations are determined by the factor $R$ defined in Eq.
(\ref{y12}). We also note that the effect of a squeezed initial
state {\it on the equilibrium part} of the correlator is to change
the effective temperature as in Eq. (\ref{y14}).

\bibliographystyle{prsty}

\begin{thebibliography}{10}

\bibitem{tang86}
M.~J. Rosker, F.~W. Wise, and C.~L. Tang, Phys. Rev. Lett. {\bf
57},  321
  (1986).

\bibitem{ruhman87}
S. Ruhman, A.~G. Joly, and K.~A. Nelson, J. Chem. Phys. {\bf 86},
6563 (1987).

\bibitem{nelson87}
Y. Yan and K.~A. Nelson, J. Chem. Phys. {\bf 87},  6240  (1987).

\bibitem{nelson87b}
Y. Yan and K. Nelson, J. Chem. Phys. {\bf 87},  6257  (1987).

\bibitem{zewail88}
A.~H. Zewail, Science {\bf 242},  1645  (1988).

\bibitem{Fragnito89}
H.~L. Fragnito, J.~Y. Bigot, P.~C. Becker, and C.~V. Shank, Chem.
Phys. Lett.
  {\bf 160},  101  (1989).

\bibitem{scherer91}
N.~F. Scherer, R. J. Carlson, A. Matro, M. Du, A. J. Ruggeiro, V.
R. Rochin, J. A. Cina, G. R. Fleming, and S. A. Rice, J. Chem.
Phys. {\bf 95},  1487 (1991).

\bibitem{pollard92}
W.~T. Pollard and R.~A. Mathies, Annu. Rev. Phys. Chem. {\bf 43},
497  (1992).

\bibitem{dexheimer92}
S.~L. Dexheimer, Q. Wang, L. A. Peteanu, W. T. Pollard, R. A.
Mathies, and C. V. Shank, Chem. Phys. Lett. {\bf 188}, 61 (1992).

\bibitem{zewail93}
A. Zewail, J. Phys. Chem. {\bf 97},  12427  (1993).

\bibitem{vos93}
M.~H. Vos, F. Rappaport, J. C. Lambry, J. Breton, and J. L.
Martin, Nature {\bf 363},  320  (1993).

\bibitem{dougherty}
T.~P. Doughery, G. P. Weiderrecht, K. A. Nelson, M. H. Garret, H.
P. Jensen, and C. Warde, Science {\bf 258}, 770 (1992).

\bibitem{nelson94}
L. Dhar, J.~A. Rogers, and K.~A. Nelson, Chem. Rev. {\bf 94},  157
(1994).

\bibitem{ZHU94}
L. Zhu, J.~T. Sage, and P.~M. Champion, Science {\bf 266},  629
(1994).

\bibitem{wang94}
Q. Wang, R. W. Schoenlein, L. A. Peteanu, R. A. Mathies, and C. V.
Shank, Science {\bf 266},  422  (1994).

\bibitem{bradforth95}
S.~E. Bradforth, T. Jiminez, F. V. Mourik, R. V. Grondelle, and G.
R. Fleming, J. Phys. Chem. {\bf 99},  16179 (1995).

\bibitem{jonas96}
D.~M. Jonas, M. J. Lang, Y. Nagasawa, T. Joo, and G. R. Fleming,
J. Phys. Chem. {\bf 100},  12660 (1996).

\bibitem{sund95}
M. Chachisvilis and V. Sundstrom, J. Chem. Phys. {\bf 104},  15
(1995).

\bibitem{sund95a}
M. Chachisvilis and V. Sundstrom, Chem. Phys. Lett. {\bf 234}, 141
(1995).

\bibitem{yang99}
T.~S. Yang, M. S. Chang, M. Hayashi, S. H. Lin, P. Vohringer, W.
Dietz, and N. F. Scherer, J. Chem. Phys. {\bf 110},  12070 (1999).

\bibitem{zhu97}
L. Zhu, A. Widom, and P.~M. Champion, J. Chem. Phys {\bf 107},
2859  (1997).

\bibitem{schuresco78}
D.~D. Schuresco and W.~W. Webb, Biophys. J. {\bf 24},  382
(1978).

\bibitem{gibson86}
Q.~H. Gibson, J.~S. Olson, R.~E. Mckinnie, and R.~J. Rohlfs, J.
Biol. Chem.
  {\bf 261},  10228  (1986).

\bibitem{miller97}
L.~M. Miller, A.~J. Pedraza, and M.~R. Chance, Biochemistry {\bf
36},  12199
  (1997).

\bibitem{hellwarth77}
R.~W. Hellwarth, Prog. Quant. Electr. {\bf 5},  1  (1977).

\bibitem{muk82}
S. Mukamel, Phys. Rep. {\bf 93},  1  (1982).

\bibitem{pollard92b}
W.~T. Pollard, S. L. Dexheimer, Q. Wang, L. A. Peteanu, C. V.
Shank, and R. A. Mathies, J. Phys. Chem. {\bf 96},  6147 (1992).

\bibitem{muk95}
S. Mukamel, {\em Principles of Nonlinear Optical Spectroscopy}
(Oxford
  University Press, New York, 1995).

\bibitem{yan89}
Y.~J. Yan, L.~E. Fried, and S. Mukamel, J. Phys. Chem. {\bf 93},
8149  (1989).

\bibitem{yan90}
Y.~J. Yan and S. Mukamel, Phys. Rev. A. {\bf 41},  6485  (1990).

\bibitem{muk90}
S. Mukamel, Ann. Rev. Phys. Chem. {\bf 41},  647  (1990).

\bibitem{fain92}
B. Fain and S. Lin, Chemical Physics {\bf 161},  515  (1992).

\bibitem{fain93b}
B. Fain, S. Lin, and V. Khidekel, Phys. Rev. A. {\bf 47},  3222
(1993).

\bibitem{fain93}
B. Fain and S. Lin, Chem. Phys. Lett. {\bf 207},  287  (1993).

\bibitem{rosca00}
F. Rosca, A. T. N. Kumar, X. Ye, T. Sjodin, A. A. Demidov, and P.
M. Champion, J. Phys. Chem. {\bf 104},  4280 (2000).

\bibitem{coalson94}
R.~D. Coalson, D.~G. Evans, and A. Nitzan, J. Chem. Phys. {\bf
101},  436
  (1994).

\bibitem{cho95}
M. Cho and R.~J. Silbey, J. Chem. Phys. {\bf 103},  595  (1995).

\bibitem{domcke97}
W. Domcke and G. Stock, Adv. Chem. Phys. {\bf 100},  1  (1997).

\bibitem{dilthey00}
S. Dilthey, S. Hahn, and G. Stock, J. Chem. Phys. {\bf 112},  4910
(2000).

\bibitem{cina93}
J.~A. Cina and T.~J. Smith, J. Chem. Phys. {\bf 98},  9211
(1993).

\bibitem{cho93a}
M. Cho, G.~R. Fleming, and S. Mukamel, J. Chem. Phys. {\bf 98},
5314  (1992).

\bibitem{cho93b}
M. Cho, M. Du, N. F. Scherer, G. R. Fleming, and S. Mukamel, J.
Chem. Phys. {\bf 99},  2410  (1993).

\bibitem{jonas95}
D.~M. Jonas, S. Bradforth, S. Passino, and G. R. Fleming, J. Phys.
Chem. {\bf 99},
   2594  (1995).

\bibitem{smith94}
T.~J. Smith, L. Ungar, and J.~A. Cina, J. Lumin. {\bf 58},  66
(1994).

\bibitem{shen99}
Y.~C. Shen and J.~A. Cina, J. Chem. Phys. {\bf 110},  9793
(1999).

\bibitem{kumar00}
A. T. N. Kumar {\it et~al.}, (in preparation).

\bibitem{sakurai}
J.~J. Sakurai, {\em Modern quantum mechanics}, revised ed.
(Addison Wesley,
  Reading MA, 1994).

\bibitem{fain90}
B. Fain and S. Lin, J. Chem. Phys. {\bf 93},  6387  (1990).

\bibitem{martin68}
P.~C. Martin, {\em Measurements and correlation functions} (Gordon
and Breach
  Science, New York, 1968).

\bibitem{banin94}
U. Banin, A. Bartana, S. Ruhman, and R. Kosloff, J. Chem. Phys.
{\bf 101},
  8461  (1994).

\bibitem{tanimura93}
Y. Tanimura and S. Mukamel, J. Opt. Soc. Am. B. {\bf 10},  2263
(1993).

\bibitem{dunn95}
T.~J. Dunn, I.~A. Walmsley, and S. Mukamel, Phys. Rev. Lett. {\bf
74},  884
  (1995).

\bibitem{foot1}
The displacement operator is the generator of coherent states of
the harmonic
  oscillator from the vaccum state. Coherent states are the nearest
  approximation to classical states of the oscillator, and derive their name
  due to their quantum statistical properties. See for example, J.~R. Klauder
  and E.~C.~G. Sudarshan, {\em Fundamentals of quantum optics} (Benjamin, New York, 1968)
\bibitem{zeigler94}
L.~D. Zeigler, R. Fan, A. Desrosiers, and N.~F. Scherer, J. Chem.
Phys. {\bf
  100},  1823  (1994).

\bibitem{constantine97}
S. Constantine, Y. Zhou, J. Morais, and L.~D. Zeigler, J. Phys.
Chem {\bf 101},
   5456  (1997).

\bibitem{page91}
J.~B. Page,  in {\em Light Scattering in Solids VI}, edited by M.
Cardona and
  G. Guntherodt (Springer, Berlin, 1991), p.\ 17.

\bibitem{gu94}
Y. Gu, A. Widom, and P.~M. Champion, J. Chem. Phys. {\bf 100},
2547  (1994).

\bibitem{foot2}
Note that the normalization factor necessary to account for the
pump induced electronic population change of the ground and
excited states, i.e. the trace of $\delta \widehat{\rho}_{g}$ and
$\delta \widehat{\rho}_{e}$ (which are unchanged in the
representation Eq. (\ref{42}) due to the unitarity of the
displacement operator) is not shown for clarity.

\bibitem{mandel95}
L. Mandel and E. Wolf, {\em Optical Coherence and Quantum Optics}
(Cambridge
  Univ. Press., Cambridge, 1995).

\bibitem{druet81}
S. Druet and J.~E. Taran, Prog. Quant. Electr. {\bf 7},  1
(1981).

\bibitem{walsh89}
A.~M. Walsh and R.~F. Loring, Chem. Phys. Lett. {\bf 160},  299
(1989).

\bibitem{foot6}
Note that the RWA employed in the derivation of Eqs. (38-44) is
still valid in the
  off-resonant region, where the real part of the lineshape dominates the
  imaginary part. However, it is straightforward to include non-RWA terms in
  the calculation, and the expressions will now also involve the lineshapes at
  negative frequencies; i.e. $\Phi_{R}(-\omega)$ and
  $\Phi_{I}(-\omega)$.

\bibitem{smith96}
T.~J. Smith and J.~A. Cina, J. Chem. Phys {\bf 104},  1272
(1996).

\bibitem{pollard90}
W. Pollard, S. Lee, and R. Mathies, J. Chem. Phys. {\bf 92},  4012
(1990).

\bibitem{dexheimer00}
S.~L. Dexheimer, A.~D.~V. Pelt, J.~A. Brozik, and B.~I. Swanson,
J. Chem. Phys.
  {\bf 104},  4308  (2000).

\bibitem{zhou99}
Y. Zhou, S. Constantine, S. Harrel, and L.~D. Zeigler, J. Chem.
Phys. {\bf
  110},  5893  (1999).

\bibitem{foot3}
For a multimode system, it is found that the pump-probe excitation
profiles for
  a given mode depend weakly on the non-equilibrium displacements of the rest
  of the optically coupled modes. This is to be contrasted with the Raman
  excitation profile, which depends only on the parameters specific to a given
  mode, with the rest of the modes in the system implicitly carried through in
  the equilibrium lineshape functions\cite{page81}.

\bibitem{page81}
J. B. Page and D. Tonks, J. Chem. Phys. {\bf 75},  5694  (1981).

\bibitem{Stallard83}
B.~R. Stallard, P.~M. Champion, P. Callis, and A.~C. Albrecht, J.
Chem. Phys.
  {\bf 78},  712  (1983).

\bibitem{Schomacker89}
K.~T. Schomacker and P.~M. Champion, J. Chem. Phys. {\bf 90},
5982  (1989).

\bibitem{srajer86}
V. Srajer, K.~T. Schomacker, and P.~M. Champion, Phys. Rev. Lett.
{\bf 57},
  1267  (1986).

\bibitem{moerner93}
W.~E. Moerner and T. Basche, Angew. Chem. {\bf 105},  537  (1993).

\bibitem{becker89}
P.~C. Becker, H. L. Fragnito, J. Y. Bigot, C. H. Brito-Crux, R. L.
Fork, and C. V. Shank, Phys. Rev. Lett. {\bf 63},  505 (1989).

\bibitem{personov}
R.~I. Personov, E.~I. Al'Shits, and L.~A. Bykovskaya, Opt. Commun.
{\bf 6},
  169  (1972).

\bibitem{champ92}
P.~M. Champion, J. Raman. Spec. {\bf 23},  557  (1992).

\bibitem{champ80}
P.~M. Champion and R. Lange, J. Chem. Phys. {\bf 73},  5947
(1980).

\bibitem{srajer91}
V. Srajer, Ph.D. thesis, Northeastern University, 1991.

\bibitem{stavrov93}
S.~S. Stavrov, Biophys. J. {\bf 65},  1942  (1993).

\bibitem{janszky94}
J. Janszky, A. Vinogradov, I. Walmsley, and J. Mostowski, Phys.
Rev. A {\bf
  50},  732  (1994).

\bibitem{demidov00}
A. Demidov et.al., (in preparation).
\bibitem{ungar97}
L.~W. Ungar and J.~A. Cina, Adv. Chem. Phys. {\bf 100},  171
(1997).

\bibitem{khidekel96}
V. Khidekel, V. Chernyak, and S. Mukamel, J. Chem. Phys. {\bf
105},  8543
  (1996).

\bibitem{foot4}
The notation for the response functions here is chosen to be
consistent with previous work \cite{muk95}. Note that the temporal
variables used here are different from those on page 213 of
reference \cite{muk95} through a simple change of variables
according to $t_{3}\rightarrow t-t_{3}, \ t_{2}\rightarrow
  t-t_{2}-t_{3}$  and $t_{1}\rightarrow t-t_{1}-t_{2}-t_{3}$.

\end{thebibliography}

\end{document}